\documentclass[11pt,a4paper]{article}
\usepackage{times}
\usepackage[pdftex]{graphicx}
\usepackage[usenames]{color}
\usepackage{cancel}
\usepackage{bm}
\usepackage{amsmath} 
\usepackage{amstext}
\usepackage[sf,bf,compact,topmarks,calcwidth,pagestyles]{titlesec}
\usepackage{titletoc}
\usepackage{fancyhdr}
\usepackage{datetime}
\usepackage[margin={1.5cm,1.5cm},font=normalsize,format=hang,labelfont=bf,textfont=it,labelsep=endash]{caption}
\usepackage{setspace}
\usepackage{enumitem}
\usepackage[style=super,nonumberlist,toc]{glossaries}

\usepackage{bibspacing}
\usepackage[compact]{titlesec}
\titlespacing{\section}{0pt}{5pt}{*0}
\titlespacing{\subsection}{0pt}{*0}{*0}
\titlespacing{\subsubsection}{0pt}{*0}{*0}

\raggedright
\definecolor{red}{rgb}{1,0,0}
\definecolor{green}{rgb}{0,1,0}
\definecolor{blue}{rgb}{0,0,1}
\definecolor{Red}{rgb}{0.8666,0.03137,0.02352}
\definecolor{Blue}{rgb}{0.00784,0.67059,0.91764}
\definecolor{Darkgreen}{rgb}{0,0.68235,0}
\definecolor{Green}{rgb}{0,0.8,0}
\definecolor{Bl}{rgb}{0,0.2,0.91764}
\definecolor{Royalblue}{rgb}{0,0.2,0.91764}
\definecolor{Brickred}{rgb}{0.644541,0.164065,0.164065}
\definecolor{Brown}{rgb}{0.6,0.4,0.4}
\definecolor{Orange}{rgb}{1,0.647059,0}
\definecolor{Indigo}{rgb}{0.746105,0,0.996109}
\definecolor{Violet}{rgb}{0.308598,0.183597,0.308598}
\definecolor{Lightgrey}{rgb}{0.762951,0.762951,0.762951}
\definecolor{Darkgrey}{rgb}{0.503548,0.503548,0.503548}
\definecolor{Pink}{rgb}{1,0.6,0.6}
\definecolor{MyLightMagenta}{cmyk}{0.1,0.8,0,0.1}
\definecolor{MyDarkBlue}{rgb}{0,0.08,0.45}

\newcommand{\Deriv}[2]{\ensuremath{\cfrac{d#1}{d#2}}}

\newcommand{\DDeriv}[2]{\ensuremath{\cfrac{d^2#1}{d#2^2}}}

\newcommand{\Veps}{\ensuremath{\varepsilon}}

\newcommand{\SfC}{\ensuremath{\boldsymbol{\mathsf{C}}}}

\newcommand{\Bnabla}{\ensuremath{\boldsymbol{\nabla}}}

\newcommand{\Bsig}{\ensuremath{\boldsymbol{\sigma}}}

\newcommand{\Bzero}{\ensuremath{\boldsymbol{0}}}

\newcommand{\Bu}{\ensuremath{\mathbf{u}}}

\newcommand{\Text}{\ensuremath{\text{ext}}}

\newcommand{\Ttop}{\ensuremath{\text{top}}}
\newcommand{\Tbot}{\ensuremath{\text{bot}}}
\newcommand{\Tcore}{\ensuremath{\text{core}}}

\newcommand{\Ttt}{\ensuremath{\text{tt}}}
\newcommand{\Ttb}{\ensuremath{\text{tb}}}
\newcommand{\Tbt}{\ensuremath{\text{bt}}}
\newcommand{\Tbb}{\ensuremath{\text{bb}}}
\newcommand{\Ttc}{\ensuremath{\text{tc}}}
\newcommand{\Tbc}{\ensuremath{\text{bc}}}
\newcommand{\Half}{\ensuremath{\frac{1}{2}}}

\newcommand{\Bdot}[2]{\ensuremath{#1\cdot#2}}

\newcommand{\Grad}[1]{\ensuremath{\Bnabla #1}}

\newcommand{\Div}[1]{\ensuremath{\Bdot{\Bnabla}{#1}}}

\newcommand{\Gradu}{\ensuremath{\Grad{\Bu}}}

\newcommand{\Partial}[2]{\ensuremath{\frac{\displaystyle\partial #1}{\displaystyle\partial #2}}}

\newcommand{\Beps}{\ensuremath{\boldsymbol{\epsilon}}}

\newcommand{\Beq}{\begin{equation}}
\newcommand{\Eeq}{\end{equation}}
\newcommand{\Bal}{\begin{aligned}}
\newcommand{\Eal}{\end{aligned}}
\newcommand{\BAl}{\begin{align}}
\newcommand{\EAl}{\end{align}}

\makeglossaries

\begin{document}
\title{The Thomsen model of inserts in sandwich composites: An evaluation}
\author{Biswajit Banerjee\thanks{Corresponding author, email: b.banerjee@irl.cri.nz} and Bryan Smith\thanks{email: bryan.smith@irl.cri.nz}\\ Industrial Research Limited \\ 24 Balfour Road, Parnell, Auckland, New Zealand}
\date{September 28, 2010}

\pagestyle{empty}
\maketitle
\pagestyle{fancy}
\setlength{\footskip}{10mm}

\setstretch{1.0} 
\begin{abstract}
   {\small
   A one-dimensional finite element model of a sandwich panel with insert 
   is derived using the approach used in the Thomsen model.  The one-dimensional model produces 
   results that are close to those of a two-dimensional axisysmmetric model. Both models assume that 
   the core is homogeneous.  Our results indicate that the one-dimensional 
   model may be well suited for small deformations of sandwich specimens with foam 
   cores.  
   }
\end{abstract}

\section{Introduction}
The numerical simulation of complicated sandwich structures containing inserts
can be computationally expensive, particularly when a statistical analysis of
the effect of variable input parameters is the goal.  Simplified theories of
sandwich structures provide a means of assessing the adequacy of the particular
statistical technique that is of interest.

Theories of sandwich structures can be broadly classified into the following 
types:
\begin{itemize}
  \item First-order theories (see for example, \cite{Plantema66}).
  \item Higher-order linear theories that do not account for thickness change (see for
        example ~\cite{Burton97} and references therein).
  \item Geometrically-exact single-layer nonlinear theories that do not account for 
        thickness change (see for example, ~\cite{VuQuoc97}).
  \item Higher-order linear single-layer theories that account for thickness change 
        (see for example, ~\cite{Anderson98, Barut01, Barut02}) .
  \item Higher-order linear multi-layer theories that account for thickness change 
        (see for example, ~\cite{Thomsen98, Thomsen98a, Thomsen00, Rabin02, Kulikov08}) .
  \item Higher-order nonlinear single-layer theories that account for thickness change
        (see for example, ~\cite{VuQuoc00,Frostig05, Arciniega07, Arciniega07a, Hohe08}).
\end{itemize}

Most theories start with ad-hoc assumptions about the displacement or stress field.
Geometrically-exact theories avoid such assumptions but are hampered by the requirement
that special constitutive models have to be designed for consistency.
The linear theory proposed by Thomsen and co-workers~\cite{Thomsen98, Thomsen98a, Thomsen00}
provides a formulation that is simple enough to be evaluated rapidly.  Therefore, we have
chosen that formulation and applied it to an axisymmetric sandwich panel in this work.
The work of Thomsen involves the solution of a system of first order ordinary differential
equations using a multi-segment numerical method,  We have instead chosen to use the 
considerably simpler finite element method to discretize and solve the system of equations.

\section{The Thomsen Model}
\begin{figure}[htb!]
  \centering
  \scalebox{0.5}{\includegraphics{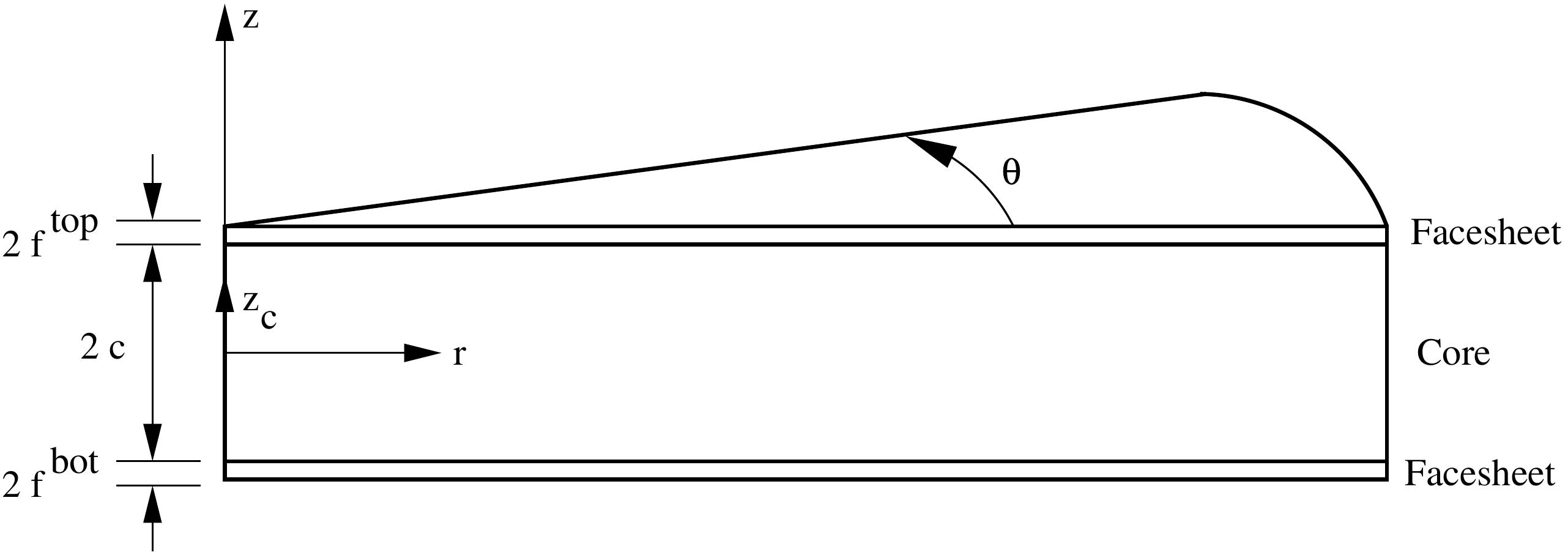}}
  \caption{The geometry of the sandwich panel.}
  \label{fig:SandwichGeom}
\end{figure}

Since we are considering a simplified axisymmetric form of the sandwich panel problem, we start 
with the governing equations expessed in cylindrical coordinates.  The geometry of the sandwich 
structure under consideration is shown in Figure~\ref{fig:SandwichGeom}.

\subsubsection{Strain-displacement}
The strain-displacement relations are given by
\Beq
  \Beps = \Half\left[\Gradu + (\Gradu)^T\right]
\Eeq
In cylindrical coordinates we have
\Beq
  \begin{aligned}
    \varepsilon_{rr} &= \Partial{u_r}{r} ~;~~
    \varepsilon_{\theta\theta} = \cfrac{1}{r}\left(\Partial{u_\theta}{\theta}+u_r\right) ~;~~
    \varepsilon_{zz} = \Partial{u_z}{z} \\
    \varepsilon_{r\theta} &= \Half\left[\Partial{u_\theta}{r} + 
         \cfrac{1}{r}\left(\Partial{u_r}{\theta}-u_\theta\right)\right] ~;~~
    \varepsilon_{\theta z} = \Half\left[\Partial{u_\theta}{z} + 
         \cfrac{1}{r}\Partial{u_z}{\theta}\right] ~;~~
    \varepsilon_{r z} = \Half\left[\Partial{u_r}{z} + \Partial{u_z}{r}\right] 
  \end{aligned}
\newglossaryentry{1}{name={$u_{r}$}, description={Displacement in the $r$-direction}}
\newglossaryentry{2}{name={$u_{\theta}$}, description={Displacement in the $\theta$-direction}}
\newglossaryentry{3}{name={$u_{z}$}, description={Displacement in the $z$-direction}}
\newglossaryentry{4}{name={$\Veps_{rr}$}, description={Strain in the $r$-direction}}
\newglossaryentry{5}{name={$\Veps_{\theta\theta}$}, description={Strain in the $\theta$-direction}}
\newglossaryentry{6}{name={$\Veps_{zz}$}, description={Strain in the $z$-direction}}
\newglossaryentry{7}{name={$\Veps_{r\theta}$}, description={Strain in the $r\theta$-plane}}
\newglossaryentry{8}{name={$\Veps_{\theta z}$}, description={Strain in the $\theta z$-plane}}
\newglossaryentry{9}{name={$\Veps_{rz}$}, description={Strain in the $rz$-plane}}
\Eeq
Axisymmetry implies that the displacement $u_\theta = u_\theta(r)$ and all derivatives with 
respect to $\theta$ are zero.  If in addition, the displacements are small such that 
$u_\theta = C~r$ (this assumption is not strictly necessary), the strain-displacement relations reduce to
\Beq
  \begin{aligned}
    \varepsilon_{rr} &= \Partial{u_r}{r} ~;~~
    \varepsilon_{\theta\theta} = \cfrac{u_r}{r} ~;~~
    \varepsilon_{zz} = \Partial{u_z}{z} \\
    \varepsilon_{\theta z} &= 0  ~;~~
    \varepsilon_{rz} = \Half\left[\Partial{u_r}{z} + \Partial{u_z}{r}\right] ~;~~
    \varepsilon_{r\theta} = 0 
  \end{aligned}
\Eeq

\subsubsection{Stress-strain}
The stress-strain relations for an orthotropic material are
\Beq
   \Bsig = \SfC:\Beps
\Eeq
In cylindrical coordinates
\Beq
  \begin{bmatrix}
    \sigma_{rr} \\ \sigma_{\theta\theta} \\ \sigma_{zz} \\ \sigma_{\theta z} 
    \\ \sigma_{rz} \\ \sigma_{r\theta} 
  \end{bmatrix} = 
  \begin{bmatrix}
    C_{11} & C_{12} & C_{13} & 0 & 0 & 0 \\
    C_{12} & C_{22} & C_{23} & 0 & 0 & 0 \\
    C_{13} & C_{23} & C_{33} & 0 & 0 & 0 \\
    0 & 0 & 0 & C_{44} & 0 & 0 \\
    0 & 0 & 0 & 0 & C_{55} & 0 \\
    0 & 0 & 0 & 0 & 0 & C_{66} 
  \end{bmatrix}
  \begin{bmatrix}
    \Veps_{rr} \\ \Veps_{\theta\theta} \\ \Veps_{zz} \\ \Veps_{\theta z} \\ 
    \Veps_{rz} \\ \Veps_{r\theta} 
  \end{bmatrix} = 
\newglossaryentry{10}{name={$\sigma_{rr}$}, description={Stress in the $r$-direction}}
\newglossaryentry{11}{name={$\sigma_{\theta\theta}$}, description={Stress in the $\theta$-direction}}
\newglossaryentry{12}{name={$\sigma_{zz}$}, description={Stress in the $z$-direction}}
\newglossaryentry{13}{name={$\sigma_{r\theta}$}, description={Stress in the $r\theta$-plane}}
\newglossaryentry{14}{name={$\sigma_{\theta z}$}, description={Stress in the $\theta z$-plane}}
\newglossaryentry{15}{name={$\sigma_{rz}$}, description={Stress in the $rz$-plane}}
\newglossaryentry{16}{name={$C_{ij}$}, description={Components of stiffness matrix}}
\Eeq
From axisymmetry, we therefore have
\Beq
   \begin{aligned}
   \sigma_{rr} & = C_{11}~\Veps_{rr} + C_{12}~\Veps_{\theta\theta} + C_{13}~\Veps_{zz} \\
   \sigma_{\theta\theta} & = C_{12}~\Veps_{rr} + C_{22}~\Veps_{\theta\theta} + C_{23}~\Veps_{zz} \\
   \sigma_{zz} & = C_{13}~\Veps_{rr} + C_{23}~\Veps_{\theta\theta} + C_{33}~\Veps_{zz} \\
   \sigma_{\theta z} & = 0 ~;~~ \sigma_{rz}  = C_{55}~\Veps_{rz} ~;~~ \sigma_{r\theta}  = 0 
   \end{aligned}
\Eeq

\subsubsection{Equilibrium}
We assume that there are no inertial or body forces in the sandwich panel.  Then the 
three-dimensional equilibrium equations take the form
\Beq
  \Div{\Bsig} = \Bzero
\Eeq
The equilibrium equations in cylindrical coordinates are
\Beq
  \begin{aligned}
    \Partial{\sigma_{rr}}{r} + \cfrac{1}{r}\left[\Partial{\sigma_{r\theta}}{\theta} + 
       (\sigma_{rr}-\sigma_{\theta\theta})\right] + \Partial{\sigma_{rz}}{z} & = 0 \\
    \Partial{\sigma_{r\theta}}{r} + \cfrac{1}{r}\left[\Partial{\sigma_{\theta\theta}}{\theta} + 
       2\sigma_{r\theta}\right] + \Partial{\sigma_{\theta z}}{z} & = 0 \\
    \Partial{\sigma_{rz}}{r} + \cfrac{1}{r}\left[\Partial{\sigma_{\theta z}}{\theta} + 
       \sigma_{rz}\right] + \Partial{\sigma_{zz}}{z} & = 0 
  \end{aligned}
\Eeq
Because of axisymmetry, all derivatives with respect to $\theta$ are zero and 
also $\sigma_{\theta z}$ and $\sigma_{r\theta}$ are zero, the reduced equilibrium equations are
\Beq
  \begin{aligned}
    \Partial{\sigma_{rr}}{r} + \cfrac{1}{r}\left[\sigma_{rr}-\sigma_{\theta\theta}\right] + 
       \Partial{\sigma_{rz}}{z} & = 0 \\
    \Partial{\sigma_{rz}}{r} + \cfrac{1}{r}\sigma_{rz} + \Partial{\sigma_{zz}}{z} & = 0 
  \end{aligned}
\Eeq

\subsection{Facesheet equations}
The facesheets are modeled using the Kirchhoff-Love hypothesis, i.e., that transverse normals
remain straight and normal and that the normals are inextensible.  In that case, the 
displacement field in the plate takes the form:
\Beq \label{eq:dispPlate}
  u_r(r,\theta,z)  = u_{0r}(r,\theta) - z~\Partial{w_0}{r} ~;~~
  u_\theta(r,\theta,z)  = u_{0\theta}(r,\theta) - z~\Partial{w_0}{\theta} ~;~~
  u_z(r,\theta,z)  = w_0(r,\theta)
\newglossaryentry{20}{name={$u_{0r}$}, description={Displacement of the plate midsurface in the $r$-direction}}
\newglossaryentry{21}{name={$u_{0\theta}$}, description={Displacement of the plate midsurface in the 
         $\theta$-direction}}
\newglossaryentry{22}{name={$w_0$}, description={Displacement of the plate midsurface in the $z$-direction}}
\Eeq
where $u_{0r}$ is the displacement of the midsurface in the $r$-direction, $u_{0\theta}$ is the
displacement of the midsurface in the $\theta$-direction, and $w_0$ is the $z$-direction
displacement of the midsurface.  

We define the stress resultants and stress couples as
\Beq \label{eq:resultantDef}
  N_{rr} := \int_{-f}^f \sigma_{rr}~dz ~;~~
  N_{\theta\theta} := \int_{-f}^f \sigma_{\theta\theta}~dz ~;~~
  M_{rr} := \int_{-f}^f z~\sigma_{rr}~dz ~;~~
  M_{\theta\theta} := \int_{-f}^f z~\sigma_{\theta\theta}~dz  
\newglossaryentry{30}{name={$f$}, description={Half the plate thickness}}
\newglossaryentry{31}{name={$N_{rr}$}, description={Stress resultant in the $r$-direction}}
\newglossaryentry{32}{name={$N_{\theta\theta}$}, description={Stress resultant in the $\theta$-direction}}
\newglossaryentry{33}{name={$M_{rr}$}, description={Moment resultant in the $r$-direction}}
\newglossaryentry{34}{name={$M_{\theta\theta}$}, description={Moment resultant in the $\theta$-direction}}
\Eeq
where the thickness of the plate is $2f$.

\subsubsection{Strain-displacement relations}
From axisymmetry, the strain-displacement relations are (for small rotations, i.e., NOT the 
von Karman strains) 
\Beq
  \varepsilon_{rr} = \Partial{u_r}{r} ~;~~
  \varepsilon_{\theta\theta} = \cfrac{u_r}{r} ~;~~
  \varepsilon_{zz} = \Partial{u_z}{z} ~;~~
  \varepsilon_{\theta z} = 0  ~;~~
  \varepsilon_{rz} = \Half\left[\Partial{u_r}{z} + \Partial{u_z}{r}\right] ~;~~
  \varepsilon_{r\theta} = 0 
\Eeq
Plugging in the displacement functions in the strain-displacement relations gives
\Beq \label{eq:plateStrainDisp}
  \begin{aligned}
    \varepsilon_{rr} &= \Deriv{u_{0r}}{r} - z~\DDeriv{w_0}{r} ~;~~
    \varepsilon_{\theta\theta} = \cfrac{u_{0r}}{r} - \cfrac{z}{r}~\Deriv{w_0}{r} ~;~~
    \varepsilon_{zz} = 0 \\
    \varepsilon_{\theta z} &= 0  ~;~~
    \varepsilon_{rz} = \Half\left[-\Deriv{w_0}{r} + \Deriv{w_0}{r}\right] = 0 ~;~~
    \varepsilon_{r\theta} = 0 
  \end{aligned}
\Eeq
To simplify the notation, we define
\Beq \label{eq:plateStrainDef}
   \begin{aligned}
     \Veps_{rr}^0(r) & := \Deriv{u_{0r}}{r} ~;~~ \Veps_{rr}^1(r)  := -\DDeriv{w_{0}}{r}  \\
     \Veps_{\theta\theta}^0(r) & := \cfrac{u_{0r}}{r} ~;~~ 
          \Veps_{\theta\theta}^1(r) := -\cfrac{1}{r}~\Deriv{w_{0}}{r} 
   \end{aligned}
\Eeq
to get
\Beq \label{eq:plateStrain}
    \varepsilon_{rr}(r,z) = \Veps_{rr}^0(r) + z~\Veps_{rr}^1(r) ~;~~
    \varepsilon_{\theta\theta}(r,z) =  \Veps_{\theta\theta}^0(r) + z~\Veps_{\theta\theta}^1(r) 
\newglossaryentry{40}{name={$\Veps_{rr}^0$}, description={Strain component in the $r$-direction independent of $w_0$}}
\newglossaryentry{41}{name={$\Veps_{rr}^1$}, description={Strain component in the $r$-direction dependent on $w_0$}}
\newglossaryentry{42}{name={$\Veps_{\theta\theta}^0$}, description={Strain component in the $\theta$-direction independent of $w_0$}}
\newglossaryentry{43}{name={$\Veps_{\theta\theta}^1$}, description={Strain component in the $\theta$-direction dependent on $w_0$}}
\Eeq

\subsubsection{Stress-strain relations}
Assuming that the facesheets are transversely isotropic and taking into account the 
strain-displacement relations (\ref{eq:plateStrainDisp}), the axisymmetric stress-strain 
relations are
\Beq
   \begin{aligned}
   \sigma_{rr} & = C_{11}~\Veps_{rr} + C_{12}~\Veps_{\theta\theta} ~;~~
   \sigma_{\theta\theta}  = C_{12}~\Veps_{rr} + C_{11}~\Veps_{\theta\theta}  ~;~~
   \sigma_{zz}  = C_{13}~\Veps_{rr} + C_{13}~\Veps_{\theta\theta} \\
   \sigma_{\theta z} & = 0 ~;~~
   \sigma_{rz}  = 0 ~;~~
   \sigma_{r\theta}  = 0 
   \end{aligned}
\Eeq
Using the definitions in (\ref{eq:plateStrainDef}) the
stress-strain relations reduce to
\Beq
   \begin{aligned}
   \sigma_{rr} & = C_{11}~\Veps_{rr}^0 + z~C_{11}~\Veps_{rr}^1 + 
         C_{12}~\Veps_{\theta\theta}^0 + z~C_{12}~\Veps_{\theta\theta}^1 \\
   \sigma_{\theta\theta} & = C_{12}~\Veps_{rr}^0 + z~C_{12}~\Veps_{rr}^1 + 
         C_{11}~\Veps_{\theta\theta}^0 + z~C_{11}~\Veps_{\theta\theta}^1 \\ 
   \sigma_{zz} & = C_{13}~\Veps_{rr}^0 + z~C_{13}~\Veps_{rr}^1 + 
         C_{13}~\Veps_{\theta\theta}^0 + z~C_{13}~\Veps_{\theta\theta}^1 
   \end{aligned}
\Eeq
If we make the plane stress assumption, $\sigma_{zz} = 0$, then we have
\Beq
   \Veps_{rr} = -\Veps_{\theta\theta} ~.
\Eeq

Then the relations between the stress resultants and stress couples and the strains are
\Beq
   \begin{aligned}
   N_{rr} & = C_{11}~\int_{-f}^f\Veps_{rr}~dz + C_{12}~\int_{-f}^f\Veps_{\theta\theta}~dz \\
   N_{\theta\theta} & = C_{12}~\int_{-f}^f\Veps_{rr}~dz + 
      C_{11}~\int_{-f}^f\Veps_{\theta\theta}~dz  \\
   M_{rr} & = C_{11}~\int_{-f}^f z\Veps_{rr}~dz + C_{12}~\int_{-f}^f z\Veps_{\theta\theta}~dz \\
   M_{\theta\theta} & = C_{12}~\int_{-f}^f z\Veps_{rr}~dz + 
      C_{11}~\int_{-f}^f z\Veps_{\theta\theta}~dz
   \end{aligned}
\Eeq
From the expressions for strain in equations (\ref{eq:plateStrain})
\Beq
  \begin{aligned}
    \int_{-f}^f \Veps_{rr}(r,z) &= 2f~\Veps_{rr}^0(r) ~;~~
    \int_{-f}^f z~\Veps_{rr}(r,z) = \cfrac{2f^3}{3}~\Veps_{rr}^1(r)\\
    \int_{-f}^f \Veps_{\theta\theta}(r,z) &=  2f~\Veps_{\theta\theta}^0(r)  ~;~~
    \int_{-f}^f z~\Veps_{\theta\theta}(r,z) =  \cfrac{2f^3}{3}~\Veps_{\theta\theta}^0(r) 
  \end{aligned}
\Eeq
Therefore, the relations between the stress resultants and stress couples and the strain can be 
expressed in matrix form as
\Beq
  \begin{bmatrix} N_{rr} \\ N_{\theta\theta} \end{bmatrix} = 
  \begin{bmatrix} A_{11} & A_{12} \\ A_{12} & A_{11} \end{bmatrix}
  \begin{bmatrix} \Veps_{rr}^0 \\ \Veps_{\theta\theta}^0 \end{bmatrix}
\Eeq
and
\Beq
  \begin{bmatrix} M_{rr} \\ M_{\theta\theta} \end{bmatrix} = 
  \begin{bmatrix} D_{11} & D_{12} \\ D_{12} & D_{11} \end{bmatrix}
  \begin{bmatrix} \Veps_{rr}^1 \\ \Veps_{\theta\theta}^1 \end{bmatrix}
\Eeq
where $A_{ij} = 2f~C_{ij}$ are the extensional stiffnesses of the plate and 
$D_{ij} = 2f^3/3~C_{ij}$ are the bending stiffnesses of the plate.
\newglossaryentry{50}{name={$A_{ij}$}, description={Extensional stiffness components of plate.}}
\newglossaryentry{51}{name={$D_{ij}$}, description={Bending stiffness components of plate.}}

\subsubsection{Equilibrium equations}
The plate equilibrium equations may be derived directly from the three-dimensional equilibrium
equations.  However, it is more informative to derive them from the principle of virtual work
\Beq
  \delta U = \delta V_{\Text}
\Eeq
where $\delta U$ is a variation of the internal energy and $\delta V_{\Text}$ is a variation of
the work done by external forces.
\newglossaryentry{60}{name={$\delta U$}, description={  Variation of the internal energy }}
\newglossaryentry{61}{name={$\delta V_{\Text}$}, description={Variation of the work done by external forces}}

The variation in the internal energy is given by
\Beq
  \delta U = \int_{\Omega_0} \int_{-f}^f \left[\sigma_{rr}~\delta\Veps_{rr} + 
    \sigma_{\theta\theta}~\delta\Veps_{\theta\theta}\right]~dz~d\Omega_0
\Eeq
where $\Omega_0$ represents the reference surface of the plate.  In terms of the definitions in
(\ref{eq:plateStrainDef}), 
\Beq
  \delta U = \int_{\Omega_0} \int_{-f}^f \left[\sigma_{rr}~\delta\Veps_{rr}^0 + 
    z~\sigma_{rr}~\delta\Veps_{rr}^1 + 
    \sigma_{\theta\theta}~\delta\Veps_{\theta\theta}^0 + 
    z~\sigma_{\theta\theta}~\delta\Veps_{\theta\theta}^1 \right]~dz~d\Omega_0
\Eeq
\newglossaryentry{70}{name={$\Omega_0$}, description={The midsurface of the plate.}}
\newglossaryentry{71}{name={$\Gamma_0$}, description={The boundary of the midsurface of the plate.}}

The definitions in (\ref{eq:resultantDef}) give
\Beq
  \delta U = \int_{\Omega_0} \left[N_{rr}~\delta\Veps_{rr}^0 + 
    M_{rr}~\delta\Veps_{rr}^1 + 
    N_{\theta\theta}~\delta\Veps_{\theta\theta}^0 + 
    M_{\theta\theta}~\delta\Veps_{\theta\theta}^1 \right]~d\Omega_0
\Eeq
Expanding out the strains in terms of the displacements, we have 
\Beq
  \delta U = \int_{\Omega_0} \left[N_{rr}~\Deriv{\delta u_{0r}}{r} - 
    M_{rr}~\DDeriv{\delta w_0}{r} + 
    \cfrac{N_{\theta\theta}}{r}~\delta u_{0r} - 
    \cfrac{M_{\theta\theta}}{r}~\Deriv{\delta w_0}{r} \right]~d\Omega_0
\Eeq
Integration by parts leads to,
\Beq
  \begin{aligned}
  \delta U & = \oint_{\Gamma_0} n_r~N_{rr}~\delta u_{0r}~d\Gamma_0 - 
      \int_{\Omega_0} \cfrac{1}{r}\Deriv{}{r}(r~N_{rr})~\delta u_{0r}~d\Omega_0   \\
      & \qquad - \oint_{\Gamma_0} n_r~M_{rr}~\Deriv{\delta w_0}{r}~d\Gamma_0 + 
      \int_{\Omega_0} \cfrac{1}{r}\Deriv{}{r}(r~M_{rr})~\Deriv{\delta w_0}{r}~d\Omega_0 +  
      \int_{\Omega_0}\cfrac{N_{\theta\theta}}{r}~\delta u_{0r}~d\Omega_0  \\
      & \qquad - \oint_{\Gamma_0} n_r~\cfrac{M_{\theta\theta}}{r}~\delta w_0~d\Gamma_0 + 
      \int_{\Omega_0}\cfrac{1}{r}~\Deriv{M_{\theta\theta}}{r}~\delta w_0~d\Omega_0
  \end{aligned}
\Eeq
keeping in mind that
\Beq
  \oint_{\Gamma_0} (\bullet)~d\Gamma_0 = \int_\theta \biggl[(\bullet)\biggr]_{r_a}^{r_b}~r~d\theta ~;~~
  \int_{\Omega_0} (\bullet)~d\Omega_0 = \int_\theta \int_r (\bullet)~r~dr~d\theta 
\Eeq
Let us define
\Beq \label{eq:betaDef}
  \beta := \Deriv{w_0}{r} ~.
\Eeq
\newglossaryentry{73}{name={$\beta$}, description={The slope of the deformed midsurface of the plate.}}

Then
\Beq
  \begin{aligned}
  \delta U & = \oint_{\Gamma_0} n_r~\left(N_{rr}~\delta u_{0r} - M_{rr}~\delta\beta 
      - \cfrac{M_{\theta\theta}}{r}~\delta w_0\right)~d\Gamma_0  \\
      & \qquad - \int_{\Omega_0} \cfrac{1}{r}\left[\left(\Deriv{}{r}(rN_{rr})
          - N_{\theta\theta}\right)~\delta u_{0r} 
      - \Deriv{}{r}(rM_{rr})~\delta \beta
      - \Deriv{M_{\theta\theta}}{r}~\delta w_0\right]~d\Omega_0
  \end{aligned}
\Eeq
To remove the derivative of $w_0$ inside the area integral we integrate again by parts to get
\Beq
  \begin{aligned}
    \delta U & = \oint_{\Gamma_0} n_r~\left[N_{rr}~\delta u_{0r} - M_{rr}~\delta\beta
            +\cfrac{1}{r}\left(\Deriv{}{r}(rM_{rr}) - M_{\theta\theta}\right)~\delta w_0
            \right]~d\Gamma_0  \\
        & \qquad - \int_{\Omega_0} \cfrac{1}{r}\left[
        \left(\Deriv{}{r}(rN_{rr}) - N_{\theta\theta}\right)~\delta u_{0r}
        + \left(\DDeriv{}{r}(rM_{rr}) 
        - \Deriv{M_{\theta\theta}}{r}\right)~\delta w_0
        \right]~d\Omega_0 
  \end{aligned}
\Eeq

The variation in the work done by the external forces is
\Beq
  \Bal
  \delta V_{\Text} & = \int_{\Omega_0} \left[
     q(r)~\delta w_0 + s(r)~(\delta u_{0r} - z_f~\delta\beta) + 
     p(r)~\delta u_{0\theta}\right]~d\Omega_0 \\
     & \qquad +
    \oint_{\Gamma_0} \int_{-f}^f \left[t_{r}~(\delta u_{0r} - z~\delta\beta) 
        + t_{\theta}~\delta u_{0\theta} + t_z~\delta w_0\right]~dz
          ~d\Gamma_0
  \Eal
\newglossaryentry{80}{name={$q(r)$}, description={Distributed force per unit area acting on the plate in the positive $z$-direction}}
\newglossaryentry{81}{name={$p(r)$}, description={Distributed force per unit area acting on the plate in the positive $r$-direction}}
\newglossaryentry{82}{name={$s(r)$}, description={Distributed force per unit area acting on the plate in the positive $\theta$-direction}}
\newglossaryentry{83}{name={$z_f$}, description={$z$-coordinate in the facesheets.}}
\newglossaryentry{84}{name={$t_r$}, description={Component of traction vector acting in the $r$-direction on the edge of the plate}}
\newglossaryentry{85}{name={$t_\theta$}, description={Component of traction vector acting in the $\theta$-direction on the edge of the plate}}
\newglossaryentry{86}{name={$t_z$}, description={Component of traction vector acting in the $z$-direction on the edge of the plate}}
\Eeq
where $q(r) = q^{\text{Top Face}}(r)+q^{\text{Bot Face}}(r)$ is a distributed surface force 
(per unit area) acting the positive $z$ direction, 
$p(r) = p^{\text{Top Face}}(r)+p^{\text{Bot Face}}(r)$ is a distributed surface force 
(per unit area) acting the positive $r$ direction, 
$s(r) = s^{\text{Top Face}}(r)+s^{\text{Bot Face}}(r)$ is a distributed surface force 
(per unit area) acting the positive $\theta$ direction, $z_f$ takes the value $+f$ at the top
of the facesheet and $-f$ at the bottom of the facesheet, and 
$\mathbf{t} = t_r~\mathbf{e}_r + t_\theta~\mathbf{e}_\theta + t_z~\mathbf{e}_z$ is the surface 
traction vector.  

\begin{figure}[htb!]
  \centering
  \scalebox{0.5}{\includegraphics{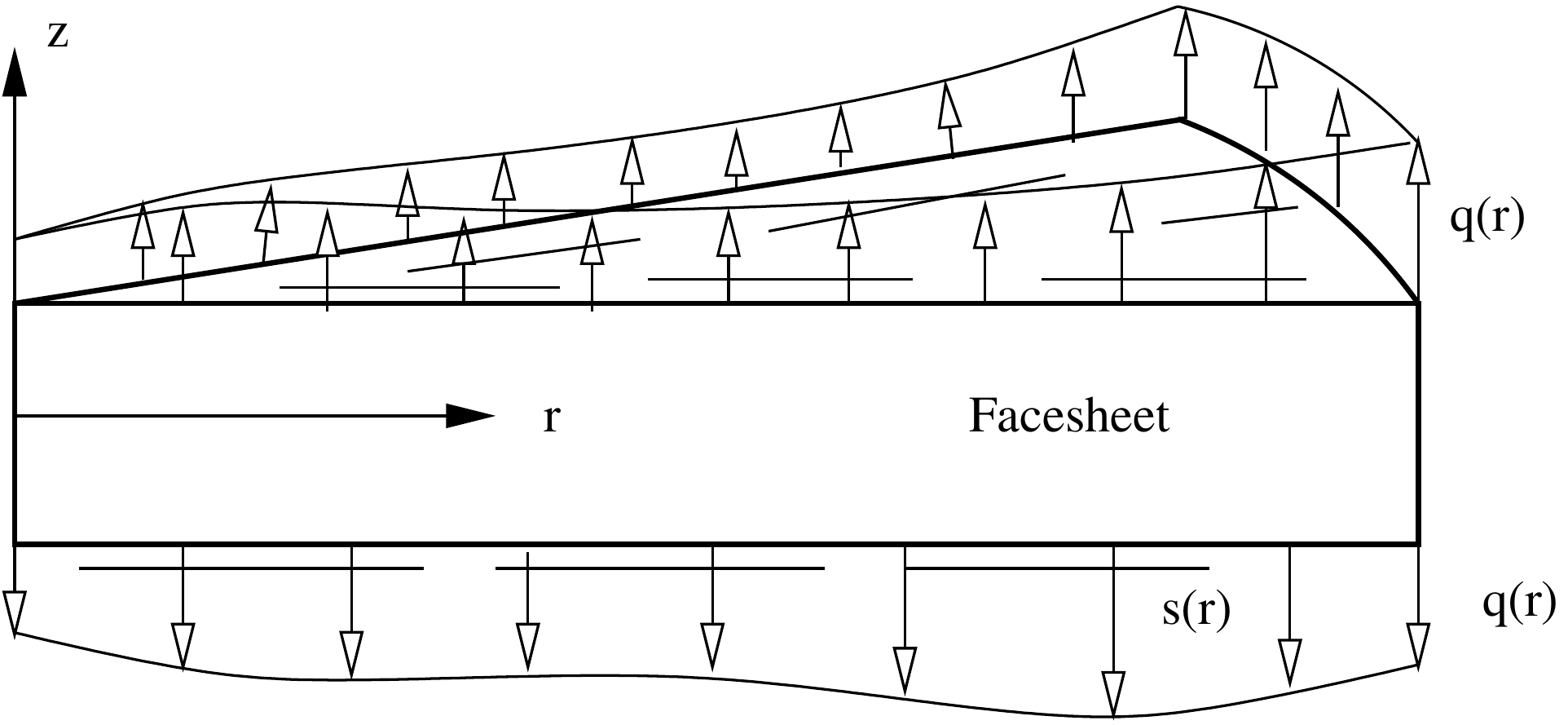}}
  \caption{The loads on a facesheet.}
  \label{fig:FaceSheetLoads}
\end{figure}

A schematic of the loads thare are applied to the facesheet is shown in Figure~\ref{fig:FaceSheetLoads}.

In terms of resultants over the thickness of the plate
\Beq
  \Bal
  \delta V_{\Text} & = \int_{\Omega_0} \left[q(r)~\delta w_0
    + s(r)~\delta u_{0r} - z_f~s(r)~\delta\beta + p(r)~\delta u_{0\theta}\right]~d\Omega_0 \\
    & \qquad + \oint_{\Gamma_0} \left[N_{r}~\delta u_{0r} - M_{r}~\delta\beta
        + N_{\theta}~\delta u_{0\theta} + Q_z~\delta w_0\right]
          ~d\Gamma_0
  \Eal
\Eeq
where
\Beq
  N_r := \int_{-f}^f t_r~dz ~;~~
  N_\theta := \int_{-f}^f t_\theta~dz ~;~~
  Q_z := \int_{-f}^f t_z~dz ~;~~
  M_r := \int_{-f}^f z~t_r~dz 
\Eeq
\newglossaryentry{90}{name={$N_r$}, description={Resultant traction in the $r$-direction of plate}}
\newglossaryentry{91}{name={$N_\theta$}, description={Resultant traction in the $\theta$-direction of plate}}
\newglossaryentry{92}{name={$Q_z$}, description={Resultant traction in the $z$-direction of plate}}
\newglossaryentry{93}{name={$M_r$}, description={Resultant moment in the $rz$-plane of plate}}

Integrating the $\delta \beta$ term by parts over the area $\Omega_0$ gives
\Beq
  \Bal
  \delta V_{\Text} & = \int_{\Omega_0} 
    \left[\left\{q(r) + \cfrac{z_f}{r}~\Deriv{}{r}~(rs)\right\}~\delta w_0
    + s(r)~\delta u_{0r} + p(r)~\delta u_{0\theta}\right]~d\Omega_0 \\
    & \qquad + \oint_{\Gamma_0} \left[N_{r}~\delta u_{0r} - M_{r}~\delta\beta
        + N_{\theta}~\delta u_{0\theta} + \left\{Q_z - n_r~z_f~s(r)\right\}~\delta w_0\right]
          ~d\Gamma_0
  \Eal
\Eeq

Then, from the principle of virtual work, we have
\Beq
  \begin{aligned}
    0 & = \oint_{\Gamma_0} \biggl[\left(n_r~N_{rr} - N_r\right)~\delta u_{0r} - 
            \left(n_r~M_{rr} - M_r\right)~\delta\beta 
            - N_\theta~\delta u_{0\theta} \\
        & \qquad \qquad \qquad
            \left.+\left\{\cfrac{n_r}{r}~\left(\Deriv{}{r}(rM_{rr}) - M_{\theta\theta}\right) 
              - Q_z + n_r~z_f~s(r)\right\}~\delta w_0 \right]~d\Gamma_0  \\
        & - \int_{\Omega_0} \left[
        \left(\cfrac{1}{r}\Deriv{}{r}(rN_{rr}) - \cfrac{N_{\theta\theta}}{r} + s(r)\right)~\delta u_{0r} 
        + p(r)~\delta u_{0\theta}\right. \\
        & \qquad \left. + \left(\cfrac{1}{r}\DDeriv{}{r}(rM_{rr}) 
            - \cfrac{1}{r}~\Deriv{M_{\theta\theta}}{r} + q(r) + \cfrac{z_f}{r}~\Deriv{}{r}~(rs)\right)
        ~\delta w_0 \right]~d\Omega_0 
  \end{aligned}
\Eeq
Because of the arbitrariness of the virtual displacements, we have
\Beq
  \begin{aligned}
    \int_{\Omega_0} \left(\cfrac{1}{r}\Deriv{}{r}(rN_{rr})
          - \cfrac{N_{\theta\theta}}{r} + s(r)\right)~\delta u_{0r}~d\Omega_0 & = 
       \oint_{\Gamma_0} (n_r~N_{rr} - N_r)~\delta u_{0r}~d\Gamma_0 \\
    \int_{\Omega_0} p(r)~\delta u_{0\theta} & = 
      -\oint_{\Gamma_0} N_{\theta}~\delta u_{0\theta}~d\Gamma_0 \\
    \int_{\Omega_0} 
      \left(\cfrac{1}{r}\DDeriv{}{r}(rM_{rr}) 
      - \cfrac{1}{r}~\Deriv{M_{\theta\theta}}{r} + q(r) + \cfrac{z_f}{r}~\Deriv{}{r}~(rs)\right)
      ~\delta w_0 & = 
       \oint_{\Gamma_0} \biggl[
         \biggl\{\cfrac{n_r}{r}~\left(\Deriv{}{r}(rM_{rr}) - M_{\theta\theta}\right)\\
         & \qquad - Q_z + n_r~z_f~s(r)\biggr\} ~\delta w_0  \\
         & \qquad \qquad - (n_r~M_{rr} - M_r)~\delta\beta\biggr] ~d\Gamma_0 
  \end{aligned}
\Eeq
Invoking the fundamental lemma of the calculus of variations (and keeping in mind that 
the displacement variations and the applied tractions are zero at points on the boundary where 
displacements are specified), we get the governing equations for the axisymmetric plate:
\Beq
  \begin{aligned}
    \cfrac{1}{r}\Deriv{(rN_{rr})}{r} - \cfrac{N_{\theta\theta}}{r} + s(r) & = 0 \\
    p(r) & = 0 \\
    \cfrac{1}{r}\DDeriv{(rM_{rr})}{r} 
        - \cfrac{1}{r}~\Deriv{M_{\theta\theta}}{r} + q(r) + \cfrac{z_f}{r}~\Deriv{}{r}~(rs) & = 0 
  \end{aligned}
\Eeq
Then the boundary conditions are 
\Beq
    \begin{aligned}
       \delta u_{0r} & : \qquad N_r = n_r~N_{rr} \\
       \delta u_{0\theta} & : \qquad  N_\theta = 0 \\
       \delta w_{0} & : \qquad Q_z = \cfrac{n_r}{r}
         \left[\Deriv{(rM_{rr})}{r} - M_{\theta\theta} + z_f~r~s(r)\right] \\
       \delta \beta & : \qquad M_r = n_r~M_{rr}
    \end{aligned}
\Eeq
The governing equations are of order 6 in the displacements $(u_{0r}, w_0)$ and there are 6 
nontrivial boundary conditions, $(u_{0r}, w_0, \partial w_0/\partial r, N_r, Q_z, M_r)$.

\subsubsection{Summary of facesheet governing equations}
The governing equations for the plate can then be summarized as follows:
\begin{itemize}
  \item Equilibrium equations:
  \Beq \label{eq:plateEq}
    \begin{aligned}
      \cfrac{1}{r}\left[\Deriv{}{r}~(rN_{rr}) - N_{\theta\theta}\right] + s(r)& = 0 \\
      \cfrac{1}{r}\left[\DDeriv{}{r}~(rM_{rr}) - \Deriv{M_{\theta\theta}}{r} 
          + z_f~\Deriv{}{r}~(rs)\right] + q(r) & = 0
    \end{aligned}
  \Eeq
  \item Stress-strain relations:
  \Beq \label{eq:plateSigEps}
    \begin{bmatrix} N_{rr} \\ N_{\theta\theta} \end{bmatrix} = 
    \begin{bmatrix} A_{11} & A_{12} \\ A_{12} & A_{11} \end{bmatrix}
    \begin{bmatrix} \Veps_{rr}^0 \\ \Veps_{\theta\theta}^0 \end{bmatrix}
  \Eeq
  \Beq
    \begin{bmatrix} M_{rr} \\ M_{\theta\theta} \end{bmatrix} = 
    \begin{bmatrix} D_{11} & D_{12} \\ D_{12} & D_{11} \end{bmatrix}
    \begin{bmatrix} \Veps_{rr}^1 \\ \Veps_{\theta\theta}^1 \end{bmatrix}
  \Eeq
  \item Strain-displacement relations:
  \Beq \label{eq:plateEpsU}
     \begin{aligned}
       \Veps_{rr}^0(r) & := \Deriv{u_{0r}}{r} ~;~~ \Veps_{rr}^1(r)  := -\DDeriv{w_{0}}{r}  \\
       \Veps_{\theta\theta}^0(r) & := \cfrac{u_{0r}}{r} ~;~~ 
            \Veps_{\theta\theta}^1(r) := -\cfrac{1}{r}~\Deriv{w_{0}}{r} 
     \end{aligned}
  \Eeq
  \item Boundary conditions:
  \Beq \label{eq:plateBC}
    \begin{aligned}
       \delta u_{0r} & : \qquad N_r = n_r~r~N_{rr} \\
       \delta w_{0} & : \qquad Q_z = \cfrac{n_r}{r}~
         \left[\Deriv{}{r}~(rM_{rr}) - M_{\theta\theta} + z_f~r~s(r)\right] \\
       \delta \beta & : \qquad M_r = n_r~r~M_{rr}
    \end{aligned}
  \Eeq
\end{itemize}

\subsubsection{Conversion into first-order ODEs}
We would like to convert the governing equations for the axisymmetric plate into ODEs of 
first order for computational purposes.  To do that, we note that the stress resultants
are related to the displacements by 
\Beq \label{eq:N_u}
  \Bal
    N_{rr} & = A_{11}~\Deriv{u_{0r}}{r} + A_{12}~\cfrac{u_{0r}}{r} \\
    N_{\theta\theta} & = A_{12}~\Deriv{u_{0r}}{r} + A_{11}~\cfrac{u_{0r}}{r} 
  \Eal
\Eeq
From the first equation in (\ref{eq:N_u}), we have
\Beq \label{eq:du0dr}
   \Deriv{u_{0r}}{r} = \cfrac{N_{rr}}{A_{11}} - \cfrac{A_{12}}{A_{11}}~\cfrac{u_{0r}}{r} ~.
\Eeq
Plugging the expression for $N_{\theta\theta}$ into the equilibrium equation for the stress
resultants (\ref{eq:plateEq}), we have
\Beq
  \cfrac{1}{r}~\Deriv{}{r}~(rN_{rr}) - \cfrac{A_{12}}{r}~\Deriv{u_{0r}}{r} - A_{11}~\cfrac{u_{0r}}{r^2} + s(r)= 0 
\Eeq
Using (\ref{eq:du0dr}), 
\Beq \label{eq:dNrrdr}
  \Deriv{N_{rr}}{r} + \left[1- \cfrac{A_{12}}{A_{11}}\right]~\cfrac{N_{rr}}{r} +
     \left(\cfrac{A_{12}^2}{A_{11}} - A_{11}\right)~\cfrac{u_{0r}}{r^2} + s(r) = 0 ~.
\Eeq
Recall that
\Beq \label{eq:dw0dr}
  \Deriv{w_0}{r} = \beta
\Eeq
Then the relations between the stress couples and the displacements take the form
\Beq \label{eq:M_beta}
  \Bal
    M_{rr} & = -D_{11}~\Deriv{\beta}{r} - D_{12}~\cfrac{\beta}{r} \\
    M_{\theta\theta} & = -D_{12}~\Deriv{\beta}{r} - D_{11}~\cfrac{\beta}{r} 
  \Eal
\Eeq
The first equation from (\ref{eq:M_beta}) can be written as
\Beq \label{eq:dbetadr}
   \Deriv{\beta}{r} = - \cfrac{M_{rr}}{D_{11}} - \cfrac{D_{12}}{D_{11}}~\cfrac{\beta}{r}
\Eeq
To convert the equilibrium equation for the stress couples into first-order ODEs, we
define
\Beq \label{eq:Qrdef}
   Q_r := \cfrac{1}{r}\Deriv{}{r}~(rM_{rr}) - \cfrac{M_{\theta\theta}}{r} + z_f~s(r)
\Eeq
\newglossaryentry{100}{name={$Q_r$}, description={Effective shear force in the $z$-direction of plate}}
Then,
\Beq \label{eq:Mrr_r_temp}
  \Deriv{M_{rr}}{r} = Q_r + \cfrac{(M_{\theta\theta}-M_{rr})}{r} - z_f~s(r)
\Eeq
Plugging in the expression for $M_{\theta\theta}$ from (\ref{eq:M_beta}) and the expression 
for the derivative of $\beta$ (\ref{eq:dbetadr}) we have
\Beq \label{eq:dMrrdr}
  \Deriv{M_{rr}}{r} = Q_r + 
     \left(\cfrac{D_{12}-D_{11}}{D_{11}}\right)~\cfrac{M_{rr}}{r} +
     \left(\cfrac{D_{12}^2-D_{11}^2}{D_{11}}\right)~\cfrac{\beta}{r^2} - z_f~s(r)
\Eeq
To reduce the order of the equilibrium equation for the stress couples, (\ref{eq:plateEq}), 
we note that taking the derivative of $Q_r$ from (\ref{eq:Qrdef}) gives us
\Beq
   r~\Deriv{Q_r}{r} + Q_r = 
    \DDeriv{}{r}~(rM_{rr}) - \Deriv{M_{\theta\theta}}{r} + z_f~\Deriv{}{r}~(rs)
\Eeq
Therefore the equilibrium equation for the stress couples can be written as
\Beq \label{eq:dQrdr}
   \Deriv{Q_{r}}{r} + \cfrac{Q_r}{r} + q(r) = 0 
\Eeq

\subsubsection{Summary first-order ODEs for facesheets}
The ODEs governing the facesheets are:
\begin{align}
   \Deriv{u_{0r}}{r} & = \cfrac{N_{rr}}{A_{11}} - \cfrac{A_{12}}{A_{11}}~\cfrac{u_{0r}}{r} \\
   \Deriv{w_0}{r} & = \beta \\
   \Deriv{\beta}{r} & = - \cfrac{M_{rr}}{D_{11}} - \cfrac{D_{12}}{D_{11}}~\cfrac{\beta}{r} \\
   \Deriv{N_{rr}}{r} & = \left[\cfrac{A_{12}-A_{11}}{A_{11}}\right]~\cfrac{N_{rr}}{r} +
     \left[\cfrac{A_{11}^2-A_{12}^2}{A_{11}}\right]~\cfrac{u_{0r}}{r^2} - s(r)\\
   \Deriv{M_{rr}}{r} & = Q_r 
     + \left[\cfrac{D_{12}-D_{11}}{D_{11}}\right]~\cfrac{M_{rr}}{r} +
     \left[\cfrac{D_{12}^2-D_{11}^2}{D_{11}}\right]~\cfrac{\beta}{r^2} - z_f~s(r)\\
   \Deriv{Q_r}{r} & = -\cfrac{Q_r}{r} - q(r) 
\end{align}
and the boundary conditions are
\Beq
  \begin{aligned}
    u_{0r} & : \qquad N_r = n_r~r~N_{rr} \\
    w_{0} & : \qquad Q_z = n_r~r~Q_r \\
    \beta & : \qquad M_r = n_r~r~M_{rr}
  \end{aligned}
\Eeq

\subsection{Core equations}
\subsubsection{Stress-strain relations}
We assume that the core is transversely isotropic.  In that case, the stress-strain relations 
in the core have the form
\Beq
   \begin{aligned}
   \sigma_{rr} & = C_{11}~\Veps_{rr} + C_{12}~\Veps_{\theta\theta} + C_{13}~\Veps_{zz} \\
   \sigma_{\theta\theta} & = C_{12}~\Veps_{rr} + C_{11}~\Veps_{\theta\theta} + C_{13}~\Veps_{zz} \\
   \sigma_{zz} & = C_{13}~\Veps_{rr} + C_{13}~\Veps_{\theta\theta} + C_{33}~\Veps_{zz} \\
   \sigma_{\theta z} & = 0  ~;~~ \sigma_{rz}  = C_{55}~\Veps_{rz} ~;~~ \sigma_{r\theta}  = 0 
   \end{aligned}
\Eeq
If we also assume that the core cannot sustain any in-plane stresses, then
\Beq
   \begin{aligned}
   \sigma_{rr}  = 0 & = C_{11}~\Veps_{rr} + C_{12}~\Veps_{\theta\theta} + C_{13}~\Veps_{zz} \\
   \sigma_{\theta\theta} = 0 & = C_{12}~\Veps_{rr} + C_{11}~\Veps_{\theta\theta} + 
      C_{13}~\Veps_{zz} 
   \end{aligned}
\Eeq
Therefore we have
\Beq
  (C_{11}-C_{12})~(\Veps_{rr}-\Veps_{\theta\theta}) = 0
\Eeq
which implies that $C_{11} = C_{12}$.  If we assume that $C_{11} = C_{12} = \epsilon~C_{13}$ 
where $\epsilon \ll 1$  is a positive quantity, then we have $C_{13} = 0$.  Therefore the 
stress-strain relations in the core reduce to
\Beq
   \sigma_{rr} = 0 ~;~~
   \sigma_{\theta\theta}  = 0 ~;~~
   \sigma_{zz}  = C_{33}~\Veps_{zz} ~;~~
   \sigma_{\theta z}  = 0 ~;~~
   \sigma_{rz}  = C_{55}~\Veps_{rz} ~;~~
   \sigma_{r\theta}  = 0 
\Eeq

\subsubsection{Strain-displacement relations}
From the strain-displacement relations we have
\Beq
    \varepsilon_{zz} = \Partial{u_z}{z} ~;~~
    \varepsilon_{rz} = \Half\left[\Partial{u_r}{z} + \Partial{u_z}{r}\right] 
\Eeq

\subsubsection{Stress-displacement relations}
Using the stress-strain relations we get
\Beq
   \sigma_{zz} = C_{33}~\Partial{u_z}{z} ~;~~
   \sigma_{rz} = \cfrac{C_{55}}{2}\left[\Partial{u_r}{z} + \Partial{u_z}{r}\right] 
\Eeq

\subsubsection{Equilibrium equations}
The equilibrium equations also reduce accordingly to
\Beq
    \Partial{\sigma_{rz}}{z}  = 0 ~;~~
    \Partial{\sigma_{rz}}{r} + \cfrac{\sigma_{rz}}{r} + \Partial{\sigma_{zz}}{z}  = 0 
\Eeq

\subsubsection{Expression for $u_z$}
Recall
\Beq
   \sigma_{zz} = C_{33}~\Partial{u_z}{z}  \qquad \implies \Partial{u_z}{z} = S_{33}~\sigma_{zz}
\Eeq
where $S_{33} := 1/C_{33}$.  
\newglossaryentry{101}{name={$S_{ij}$}, description={$1/C_{ij}$ - not components of the compliance tensor.}}

Integrating, we get
\Beq
  u_z(r,z) = \int_{z_a}^{z_b} S_{33}~\sigma_{zz}~dz + A(r)
\Eeq
where $A(r)$ is a function only of $r$.  Integrating by parts, we have
\Beq
  u_z(r,z) = S_{33}\left[\left.z~\sigma_{zz}\right|_{z_a}^{z_b} - 
         \int_{z_a}^{z_b} z~\Partial{\sigma_{zz}}{z}~dz\right] + A(r) 
\Eeq
\newglossaryentry{102}{name={$z_a-z_b$}, description={Integration limits}}
\newglossaryentry{103}{name={$z_c$}, description={z-coordinate in the core with origin at the core midsurface}}
\newglossaryentry{104}{name={$c$}, description={Half the core thickness}}

Now we assume that the displacement $u_z$ is quadratic in $z$ to get
\Beq \label{eq:Br_orig}
  \Partial{\sigma_{zz}}{z} = - \Partial{\sigma_{rz}}{r} - \cfrac{\sigma_{rz}}{r} =: B(r)
\Eeq
where $B(r)$ is a function only of $r$.  If we set up the coordinate system in the core such 
that $z_c = z - c$ where $2c$ is the core thickness and integrate from $0$ to $z_c$, we get
\Beq
  \begin{aligned}
  u_z(r,z_c) & = S_{33}\left[z_c~\sigma_{zz}(r, z_c) - B(r)~\int_{z_c}^{0} z~dz\right] + A(r)  \\
    & = S_{33}\left[z_c~\sigma_{zz}(r, z_c) - B(r)~\cfrac{z_c^2}{2}\right] + A(r)  
  \end{aligned}
\Eeq
At $z_c = c$ the displacement of the core is equal to the displacement of the top facesheet, 
i.e.,
\Beq
  w^1(r) = u_z(r,c) = S_{33}\left[c~\sigma_{zz}(r, c) - B(r)~\cfrac{c^2}{2}\right] + A(r)  
\Eeq
\newglossaryentry{201}{name={$w^1$}, description={$z$-displacement at the bottom of the top facesheet}}
\newglossaryentry{202}{name={$w^2$}, description={$z$-displacement at the top of the bottom facesheet}}
\newglossaryentry{203}{name={$A(r)$}, description={Constant of integration}}
\newglossaryentry{204}{name={$B(r)$}, description={$\partial\sigma_{zz}/\partial z$ in the core}}
Eliminating $A(r)$, we get
\Beq \label{eq:uz_zc0}
  u_z(r,z_c) = w^1(r) + S_{33}\left[\left\{z_c~\sigma_{zz}(r, z_c) - c~\sigma_{zz}(r,c)\right\} - 
         \cfrac{B(r)}{2}~\left(z_c^2-c^2\right)\right]  ~.
\Eeq
We can also calculate the displacement at the bottom facesheet
\Beq
  w^2(r) = u_z(r,-c) = S_{33}\left[-c~\sigma_{zz}(r, -c) - B(r)~\cfrac{c^2}{2}\right] + A(r)  
\Eeq
Again, eliminating $A(r)$, we have
\Beq \label{eq:w1_w2}
  w^1(r)-w^2(r) = c~S_{33}\left[\sigma_{zz}(r,c)+\sigma_{zz}(r,-c)\right] ~.
\Eeq

\subsubsection{Eliminating $\sigma_{zz}$}
We would like to eliminate $\sigma_{zz}$ from the expression in equation (\ref{eq:uz_zc_temp1}).  
To do that, we recall that
\Beq
  \Partial{\sigma_{zz}}{z} = B(r)
\Eeq
Integrating between the limits $0$ and $z_c$ as before, we get
\Beq
  \sigma_{zz}(r,z_c) = B(r)~z_c + E(r)
\Eeq
where $E(r)$ is a function of $r$ only.  
\newglossaryentry{205}{name={$E(r)$}, description={Constant of integration}}
Therefore,
\Beq
  \sigma_{zz}(r,c) = B(r)~c + E(r) ~;~~
  \sigma_{zz}(r,-c)  = -B(r)~c + E(r) 
\Eeq
which gives
\Beq
  E(r) = \sigma_{zz}(r,c) - c~B(r)~.
\Eeq
Therefore, 
\Beq \label{eq:sigzz_zc0}
  \sigma_{zz}(r,z_c) = (z_c-c)~B(r) + \sigma_{zz}(r,c)
\Eeq
We also have,
\Beq
  \sigma_{zz}(r,c)+\sigma_{zz}(r,-c) = 2E(r) = 2[\sigma_{zz}(r,c) - c~B(r)]
\Eeq
Hence, from (\ref{eq:w1_w2}),
\Beq
  w^1(r)-w^2(r) = 2~c~S_{33}~[\sigma_{zz}(r,c) - c~B(r)]
\Eeq
or,
\Beq \label{eq:sigzz_c}
  \sigma_{zz}(r,c) = \cfrac{C_{33}}{2c}~\left[w^1(r)-w^2(r)\right] + c~B(r)
\Eeq
Combining (\ref{eq:sigzz_zc0}) and (\ref{eq:sigzz_c}),
\Beq \label{eq:sigzz_zc}
  \sigma_{zz}(r,z_c) =  \cfrac{C_{33}}{2c}~\left[w^1(r)-w^2(r)\right] + z_c~B(r)
\Eeq
Using (\ref{eq:sigzz_c}) and (\ref{eq:sigzz_zc}) in (\ref{eq:uz_zc0}) gives
\Beq \label{eq:uz_zc_temp1}
  u_z(r,z_c) = w^1(r) + \left(\cfrac{z_c-c}{2}\right)\left[\cfrac{w^1(r)-w^2(r)}{c} + 
       (z_c+c)~S_{33}~B(r)\right] ~.
\Eeq
Now, from equations (\ref{eq:dispPlate}) for the facesheets, we have
\Beq
   w^1(r) = w_{0}^{\Ttop} ~;~~ w^2(r) = w_{0}^{\Tbot} 
\Eeq
respectively.  Plugging these into (\ref{eq:uz_zc_temp1}) gives
\Beq \label{eq:uz_zc}
  u_z(r,z_c) = \Half\left(\cfrac{z_c}{c}+1\right)~w_0^{\Ttop} - 
               \Half\left(\cfrac{z_c}{c}-1\right)~w_0^{\Tbot} +
               \Half\left(z_c^2-c^2\right)~S_{33}~B(r) ~.
\Eeq
\newglossaryentry{210}{name={$w_0^{\Ttop}$}, description={$z$-displacement at the midsurface of the top facesheet}}
\newglossaryentry{211}{name={$w_0^{\Tbot}$}, description={$z$-displacement at the midsurface of the bottom facesheet}}

\subsubsection{Expression for $u_r$}
Recall that
\Beq
   \sigma_{rz}  = \cfrac{C_{55}}{2}\left[\Partial{u_r}{z} + \Partial{u_z}{r}\right] 
\Eeq
Therefore,
\Beq \label{eq:dur_dz_1}
  \Partial{u_r}{z} =  2~S_{55}~\sigma_{rz} - \Partial{u_z}{r} ~;\qquad S_{55} := 1/C_{55}
\Eeq
Also, taking the $r$-derivative of equation (\ref{eq:uz_zc}), we have
\Beq \label{eq:duz_dr_1}
  \Partial{u_z}{r} = \Half\left(\cfrac{z_c}{c}+1\right)~\Partial{w_0^{\Ttop}}{r} - 
               \Half\left(\cfrac{z_c}{c}-1\right)~\Partial{w_0^{\Tbot}}{r} +
               \Half\left(z_c^2-c^2\right)~S_{33}~\Partial{B}{r} ~.
\Eeq
Substitution of (\ref{eq:duz_dr_1}) into (\ref{eq:dur_dz_1}) gives
\Beq \label{eq:dur_dz_2}
  \Partial{u_r}{z} =  2~S_{55}~\sigma_{rz} 
   - \Half\left(\cfrac{z_c}{c}+1\right)~\Deriv{w_0^{\Ttop}}{r} 
   + \Half\left(\cfrac{z_c}{c}-1\right)~\Deriv{w_0^{\Tbot}}{r} 
   - \Half\left(z_c^2-c^2\right)~S_{33}~\Deriv{B}{r} ~.
\Eeq
Note that $\Partial{\sigma_{rz}}{z} = 0 \implies \sigma_{rz} = \sigma_{rz}(r)$. 
Integrating (\ref{eq:dur_dz_2}) between $0$ and $z_c$, we get
\Beq \label{eq:ur_1}
  \begin{aligned}
  u_r(r, z_c) =  2~S_{55}~z_c~\sigma_{rz}
   - \Half\left(\cfrac{z^2_c}{2c}+z_c\right)~\Deriv{w_0^{\Ttop}}{r} 
   + \Half\left(\cfrac{z^2_c}{2c}-z_c\right)~\Deriv{w_0^{\Tbot}}{r} \\
   - \Half\left(\cfrac{z_c^3}{3}-c^2~z_c\right)~S_{33}~\Deriv{B}{r} + G(r)
  \end{aligned}
\Eeq
where $G(r)$ is a function only of $r$.
\newglossaryentry{212}{name={$G(r)$}, description={Constant of integration}}

At $z_c = c$, $u_r = u^{1}(r)$.  Hence we have
\Beq \label{eq:Gr}
  G(r) = u^1 - 2~S_{55}~c~\sigma_{rz} 
   + \cfrac{3c}{4}~\Deriv{w_0^{\Ttop}}{r} 
   + \cfrac{c}{4}~\Deriv{w_0^{\Tbot}}{r} 
   - \cfrac{c^3}{3}~S_{33}~\Deriv{B}{r} 
\Eeq
Substitution of (\ref{eq:Gr}) into (\ref{eq:ur_1}) gives
\Beq \label{eq:urCore_temp1}
  \begin{aligned}
  u_r(r, z_c) =  u^1 + 2~S_{55}~(z_c-c)~\sigma_{rz} 
   + \left[\cfrac{3c}{4} - \Half\left(\cfrac{z^2_c}{2c}+z_c\right)\right]~\Deriv{w_0^{\Ttop}}{r} 
   + \left[\cfrac{c}{4} + \Half\left(\cfrac{z^2_c}{2c}-z_c\right)\right]~\Deriv{w_0^{\Tbot}}{r} \\
   - \left[\cfrac{c^3}{3} + \Half\left(\cfrac{z_c^3}{3}-c^2~z_c\right)\right]~S_{33}~\Deriv{B}{r}
  \end{aligned}
\Eeq
Now, from equations (\ref{eq:dispPlate}) and (\ref{eq:betaDef}) for the facesheets, we have
\Beq
   \Bal
   \Deriv{w_0^\Ttop}{r} & =: \beta^{\Ttop} ~;~~ \Deriv{w_0^\Tbot}{r} =: \beta^{\Tbot} \\
   u^1(r) & = u_{0r}^{\Ttop} + f^{\Ttop}~\beta^{\Ttop} ~;~~
   u^2(r)  = u_{0r}^{\Tbot} - f^{\Tbot}~\beta^{\Tbot} 
   \Eal
\Eeq
where $2~f^{\Ttop}$ and $2~f^{\Tbot}$ are the thicknesses of the top and bottom facesheets,
respectively.  Plugging these into (\ref{eq:urCore_temp1} gives
\Beq \label{eq:urCore_temp2}
  \begin{aligned}
  u_r(r, z_c) =  u_{0r}^{\Ttop} + f^{\Ttop}~\beta^{\Ttop} + 2~S_{55}~(z_c-c)~\sigma_{rz} 
   + \left[\cfrac{3c}{4} - \Half\left(\cfrac{z^2_c}{2c}+z_c\right)\right]~\beta^{\Ttop} \\
   + \left[\cfrac{c}{4} + \Half\left(\cfrac{z^2_c}{2c}-z_c\right)\right]~\beta^{\Tbot} 
   - \left[\cfrac{c^3}{3} + \Half\left(\cfrac{z_c^3}{3}-c^2~z_c\right)\right]~S_{33}~\Deriv{B}{r}
  \end{aligned}
\Eeq
\newglossaryentry{220}{name={$f^{\Ttop}$}, description={Half the thickness of the top facesheet}}
\newglossaryentry{221}{name={$f^{\Tbot}$}, description={Half the thickness of the bottom facesheet}}
or,
\Beq \label{eq:urCore}
  \begin{aligned}
  u_r(r, z_c) =  u_{0r}^{\Ttop} 
   + \left[f^\Ttop + \cfrac{3c}{4} - \Half\left(\cfrac{z^2_c}{2c}+z_c\right)\right]~\beta^{\Ttop} 
   + \left[\cfrac{c}{4} + \Half\left(\cfrac{z^2_c}{2c}-z_c\right)\right]~\beta^{\Tbot} \\
   + 2~S_{55}~(z_c-c)~\sigma_{rz} 
   - \left[\cfrac{c^3}{3} + \Half\left(\cfrac{z_c^3}{3}-c^2~z_c\right)\right]~S_{33}~\Deriv{B}{r}
  \end{aligned}
\Eeq

\subsubsection{Governing equation for the core}
Now, at the bottom of the core, $z_c = -c$. From (\ref{eq:urCore}) we have
\Beq \label{eq:urCore_temp3}
  u_r(r,-c) = u^2 =  u^{\Ttop}_{0r} + \left[f^{\Ttop}+c\right]~\beta^{\Ttop}  
   + c~\beta^{\Tbot} - 4~S_{55}~\sigma_{rz}~c - \cfrac{2}{3}~S_{33}~c^3 ~\Deriv{B(r)}{r} 
\Eeq
Also
\[
   u^2  = u_{0r}^{\Tbot} - f^{\Tbot}~\beta^{\Tbot} 
\]
Hence
\Beq \label{eq:urCore_temp4}
   0 =  u^{\Ttop}_{0r} - u_{0r}^{\Tbot} + \left[f^{\Ttop}+c\right]~\beta^{\Ttop}  
   + \left[f^{\Tbot} + c\right]~\beta^{\Tbot} 
   - 4~S_{55}~\sigma_{rz}~c - \cfrac{2}{3}~S_{33}~c^3 ~\Deriv{B(r)}{r} 
\Eeq
Recall that
\[
   B = -\Deriv{\sigma_{rz}}{r} - \cfrac{\sigma_{rz}}{r}
\]
Therefore,
\Beq \label{eq:dBdr}
   \Deriv{B}{r} = -\DDeriv{\sigma_{rz}}{r} + \cfrac{\sigma_{rz}}{r^2}
     - \cfrac{1}{r}~\Deriv{\sigma_{rz}}{r}
\Eeq
Plugging (\ref{eq:dBdr}) into (\ref{eq:urCore_temp4}) gives
\Beq \label{eq:urCore_temp5}
   \Bal
   0 & =  u^{\Ttop}_{0r} - u_{0r}^{\Tbot} + \left[f^{\Ttop}+c\right]~\beta^{\Ttop}  
   + \left[f^{\Tbot} + c\right]~\beta^{\Tbot} \\
   & \qquad - 4~S_{55}~\sigma_{rz}~c - \cfrac{2}{3}~S_{33}~c^3\left[
   -\DDeriv{\sigma_{rz}}{r} + \cfrac{\sigma_{rz}}{r^2} - \cfrac{1}{r}~\Deriv{\sigma_{rz}}{r}
    \right]
   \Eal
\Eeq
or,
\Beq \label{eq:ddsigrz_dr}
   \Bal
   \DDeriv{\sigma_{rz}}{r} & + \cfrac{1}{r}~\Deriv{\sigma_{rz}}{r}
     - \left(\cfrac{1}{r^2} + \cfrac{6C_{33}S_{55}}{c^2}\right)~\sigma_{rz} \\
     & \qquad = -\cfrac{3C_{33}}{2c^3}~\left[u^{\Ttop}_{0r} - u_{0r}^{\Tbot} + 
        \left(f^{\Ttop}+c\right)~\beta^{\Ttop} + \left(f^{\Tbot} + c\right)~\beta^{\Tbot} \right]
   \Eal
\Eeq

\subsubsection{Conversion into first order ODEs}
To convert (\ref{eq:ddsigrz_dr}) into first-order ODEs, we
define
\Beq
   T_r := \Deriv{\sigma_{rz}}{r}
\Eeq
\newglossaryentry{222}{name={$T_r$}, description={$d\sigma_{rz}/dr$ in the core}}
Then equation (\ref{eq:ddsigrz_dr}) can be written as
\Beq \label{eq:dTr_dr}
  \Bal
   \Deriv{T_r}{r} & =  
     -\cfrac{3C_{33}}{2c^3}~\left[u^{\Ttop}_{0r} - u_{0r}^{\Tbot} + 
        \left(f^{\Ttop}+c\right)~\beta^{\Ttop} + \left(f^{\Tbot} + c\right)~\beta^{\Tbot} \right]
     \\
     & \qquad 
     + \left(\cfrac{1}{r^2} + \cfrac{6C_{33}S_{55}}{c^2}\right)~\sigma_{rz} - \cfrac{T_r}{r}
  \Eal
\Eeq

\subsubsection{Summary of first order ODEs for the core}
The governing equations for the stresses in the core are
\begin{align}
   \Deriv{\sigma_{rz}}{r} & = T_r \\
   \Deriv{T_r}{r} & =  
     -\cfrac{3C_{33}}{2c^3}~\left[u^{\Ttop}_{0r} - u_{0r}^{\Tbot} + 
        \left(f^{\Ttop}+c\right)~\beta^{\Ttop} + \left(f^{\Tbot} + c\right)~\beta^{\Tbot} \right]
       \nonumber \\
     & \qquad 
     + \left(\cfrac{1}{r^2} + \cfrac{6C_{33}S_{55}}{c^2}\right)~\sigma_{rz} - \cfrac{T_r}{r}\\
  \sigma_{zz}(r,z_c) & =  \cfrac{C_{33}}{2c}~\left[w^1(r)-w^2(r)\right] -z_c~T_r 
     - z_c~\cfrac{\sigma_{rz}}{r}
\end{align}

\section{Coupled governing equations of the facesheets and the core}
In the previous section, ODEs have been derived that partially couple the core to the facesheets.  
To complete the coupling of the facesheets to the core we have to balance the forces at the 
interfaces between the core and the facesheets.  We introduce some new notation to aid us
in the coupling process.  Recall that for a facesheet
\Beq
   s(r) = s^{\text{Top Face}} + s^{\text{Bot Face}} ~;~~
   q(r) = q^{\text{Top Face}} + q^{\text{Bot Face}} 
\Eeq
We identify these two sets of applied tractions on the two facesheets using the notation
\Beq
   s^\Ttop(r) = s^{\Ttt} + s^{\Ttb} ~;~~
   q^\Ttop(r) = q^{\Ttt} + q^{\Ttb} ~;~~
   s^\Tbot(r) = s^{\Tbt} + s^{\Tbb} ~;~~
   q^\Tbot(r) = q^{\Tbt} + q^{\Tbb} 
  \newglossaryentry{230}{name={$s^\Ttop$}, description={Total applied shear traction on top face sheet}}
  \newglossaryentry{231}{name={$s^\Ttt$}, description={Applied shear traction on top face of top face sheet}}
  \newglossaryentry{232}{name={$s^\Ttb$}, description={Applied shear traction on bottom face of top face sheet}}
  \newglossaryentry{233}{name={$s^\Tbot$}, description={Total applied shear traction on bottom face sheet}}
  \newglossaryentry{234}{name={$s^\Tbt$}, description={Applied shear traction on top face of bottom face sheet}}
  \newglossaryentry{235}{name={$s^\Tbb$}, description={Applied shear traction on bottom face of bottom face sheet}}
  \newglossaryentry{236}{name={$q^\Ttop$}, description={Total applied normal traction on top face sheet}}
  \newglossaryentry{237}{name={$q^\Ttt$}, description={Applied normal traction on top face of top face sheet}}
  \newglossaryentry{238}{name={$q^\Ttb$}, description={Applied normal traction on bottom face of top face sheet}}
  \newglossaryentry{239}{name={$q^\Tbot$}, description={Total applied normal traction on bottom face sheet}}
  \newglossaryentry{240}{name={$q^\Tbt$}, description={Applied normal traction on top face of bottom face sheet}}
  \newglossaryentry{241}{name={$q^\Tbb$}, description={Applied normal traction on bottom face of bottom face sheet}}
\Eeq
The tractions at the core-facesheet interface are given by
\Beq
  \begin{aligned}
  \mathbf{t} & = t_r~\mathbf{e}_r + t_\theta~\mathbf{e}_\theta + t_z~\mathbf{e}_z\\
    & = (n_r~\sigma_{rr} + n_\theta~\sigma_{r\theta} + n_z~\sigma_{rz})~\mathbf{e}_r + 
        (n_r~\sigma_{r\theta} + n_\theta~\sigma_{\theta\theta} + n_z~\sigma_{\theta z})~
        \mathbf{e}_\theta \\
    & \qquad + (n_r~\sigma_{rz} + n_\theta~\sigma_{\theta z} + n_z~\sigma_{zz})~\mathbf{e}_z 
  \end{aligned}
\Eeq
where $\mathbf{e}_r, \mathbf{e}_\theta, \mathbf{e}_z$ are the basis vectors in the 
$r, \theta, z$ directions.  In the core $\sigma_{rr}=\sigma_{\theta\theta} = \sigma_{\theta z}
= \sigma_{r \theta} = 0$.  Therefore, the traction vector simplifies to
\Beq
  \mathbf{t} = n_z~\sigma_{rz}~\mathbf{e}_r + (n_r~\sigma_{rz} + n_z~\sigma_{zz})~\mathbf{e}_z 
  \newglossaryentry{250}{name={$\mathbf{e}_r$}, description={Basis vector in the $r$-direction}}
  \newglossaryentry{251}{name={$\mathbf{e}_\theta$}, description={Basis vector in the $\theta$-direction}}
  \newglossaryentry{252}{name={$\mathbf{e}_z$}, description={Basis vector in the $z$-direction}}
  \newglossaryentry{253}{name={$n_r$}, description={$r$-component of surface normal vector}}
  \newglossaryentry{254}{name={$n_\theta$}, description={$\theta$-component of surface normal vector}}
  \newglossaryentry{255}{name={$n_z$}, description={$z$-component of surface normal vector}}
\Eeq
At the interface between the core and the top facesheet, $n_r = 0, n_z = 1$ while at the interface
between the core and the bottom facesheet $n_r = 0, n_z = -1$.  Therefore, 
\Beq
  \mathbf{t}^{\Ttc}(r) = \sigma_{rz}^\Tcore(r)~\mathbf{e}_r + \sigma_{zz}^\Tcore(r,c)~\mathbf{e}_z ~;~~
  \mathbf{t}^{\Tbc}(r) = -\sigma_{rz}^\Tcore(r)~\mathbf{e}_r - \sigma_{zz}^\Tcore(r,-c)~\mathbf{e}_z 
  \newglossaryentry{260}{name={$\mathbf{t}^\Ttc$}, description={Traction at interface of core and top facesheet}}
  \newglossaryentry{261}{name={$\mathbf{t}^\Tbc$}, description={Traction at interface of core and bottom facesheet}}
\Eeq
To couple the facesheet equations to the core equations we have, due to the continuity
of tractions at the core-facesheet interfaces, 
\Beq
  \Bal
  s^{\Ttb}(r) & + \mathbf{t}^\Ttc(r) \cdot \mathbf{e}_r = 0 \qquad \implies \qquad
  s^{\Ttb}(r) = -\sigma_{rz}^\Tcore(r) \\
  s^{\Tbt}(r) & + \mathbf{t}^\Tbc(r) \cdot \mathbf{e}_r = 0 \qquad \implies \qquad
  s^{\Tbt}(r) = \sigma_{rz}^\Tcore(r) \\
  q^{\Ttb}(r) & + \mathbf{t}^\Ttc(r) \cdot \mathbf{e}_z = 0 \qquad \implies \qquad
  q^{\Ttb}(r) = -\sigma_{zz}^\Tcore(r,c) \\
  q^{\Tbt}(r) & + \mathbf{t}^\Tbc(r) \cdot \mathbf{e}_z = 0 \qquad \implies \qquad
  q^{\Tbt}(r) = \sigma_{zz}^\Tcore(r,-c) 
  \Eal
\Eeq
From equations (\ref{eq:sigzz_zc}) and (\ref{eq:Br_orig})
\Beq
  \sigma^\Tcore_{zz}(r,z_c) =  \cfrac{C_{33}}{2c}~\left[w_0^\Ttop-w_0^\Tbot\right] 
   - z_c~\Deriv{\sigma_{rz}^\Tcore}{r} - z_c~\cfrac{\sigma_{rz}^\Tcore}{r}
\Eeq
Therefore,
\Beq
  \Bal
  q^{\Ttb}(r) & = -\cfrac{C_{33}^\Tcore}{2c}~\left[w_0^\Ttop-w_0^\Tbot\right] 
   + c~\Deriv{\sigma_{rz}^\Tcore}{r} + c~\cfrac{\sigma_{rz}^\Tcore}{r} \\
  q^{\Tbt}(r) & = \cfrac{C_{33}^\Tcore}{2c}~\left[w_0^\Ttop-w_0^\Tbot\right] 
   + c~\Deriv{\sigma_{rz}^\Tcore}{r} + c~\cfrac{\sigma_{rz}^\Tcore}{r} 
  \Eal
\Eeq
Equation (\ref{eq:dQrdr}) then takes the form
\Beq
  \Bal
  \Deriv{Q^{\Ttop}_r}{r} = -\cfrac{Q_r^\Ttop}{r} -q^{\Ttop}(r) & = 
    - \cfrac{Q_r^\Ttop}{r} 
    + \cfrac{C_{33}^\Tcore}{2c}~\left[w_0^\Ttop-w_0^\Tbot\right] 
    - c~\Deriv{\sigma_{rz}^\Tcore}{r} - c~\cfrac{\sigma_{rz}^\Tcore}{r}
    - q^\Ttt \\
  \Deriv{Q^{\Tbot}_r}{r} = -\cfrac{Q_r^\Tbot}{r} -q^{\Tbot}(r) & = 
    - \cfrac{Q_r^\Tbot}{r} 
    - \cfrac{C_{33}^\Tcore}{2c}~\left[w_0^\Ttop-w_0^\Tbot\right] 
    - c~\Deriv{\sigma_{rz}^\Tcore}{r} - c~\cfrac{\sigma_{rz}^\Tcore}{r} 
    - q^\Tbb
  \Eal
\Eeq
Similarly, equation (\ref{eq:dNrrdr}) takes the form
\Beq
  \Deriv{N^\Ttop_{rr}}{r} = 
   \left[\cfrac{A^\Ttop_{12}-A^\Ttop_{11}}{A^\Ttop_{11}}\right]~\cfrac{N^\Ttop_{rr}}{r} +
   \left[\cfrac{(A^\Ttop_{11})^2 - (A^\Ttop_{12})^2}{A^\Ttop_{11}}\right]~\cfrac{u^\Ttop_{0r}}{r^2} 
   + \sigma^\Tcore_{rz} - s^{\Ttt}
\Eeq
and
\Beq
  \Deriv{N^\Tbot_{rr}}{r} = 
   \left[\cfrac{A^\Tbot_{12}-A^\Tbot_{11}}{A^\Tbot_{11}}\right]~\cfrac{N^\Tbot_{rr}}{r} +
   \left[\cfrac{(A^\Tbot_{11})^2 - (A^\Tbot_{12})^2}{A^\Tbot_{11}}\right]~\cfrac{u^\Tbot_{0r}}{r^2} 
   - \sigma^\Tcore_{rz} - s^{\Tbb}
\Eeq
Also, equation (\ref{eq:dMrrdr}) takes the form
\Beq
  \Deriv{M^\Ttop_{rr}}{r} = Q^\Ttop_r
   + \left[\cfrac{D^\Ttop_{12}-D^{\Ttop}_{11}}{D^\Ttop_{11}}\right]~\cfrac{M^\Ttop_{rr}}{r} +
     \left[\cfrac{(D^\Ttop_{12})^2-(D^\Ttop_{11})^2}{D^\Ttop_{11}}\right]~\cfrac{\beta^\Ttop}{r^2} 
     + f^\Ttop~(s^\Ttt-\sigma^\Tcore_{rz})
\Eeq
and
\Beq
  \Deriv{M^\Tbot_{rr}}{r} = Q^\Tbot_r 
    + \left[\cfrac{D^\Tbot_{12}-D^{\Tbot}_{11}}{D^\Tbot_{11}}\right]~\cfrac{M^\Tbot_{rr}}{r} +
     \left[\cfrac{(D^\Tbot_{12})^2-(D^\Tbot_{11})^2}{D^\Tbot_{11}}\right]~\cfrac{\beta^\Tbot}{r^2} 
     - f^\Tbot~(s^\Tbb+\sigma^\Tcore_{rz})
\Eeq

The governing first order ODEs for the facesheets and the core can then be expressed as

\begin{itemize}
   \item {\bf Top facesheet}:
   \Beq
     \begin{aligned}
   \Deriv{u^{\Ttop}_{0r}}{r} & = \cfrac{N_{rr}^{\Ttop}}{A_{11}^{\Ttop}} - \cfrac{A_{12}^{\Ttop}}{A_{11}^{\Ttop}}~\cfrac{u_{0r}^{\Ttop}}{r} \\
   \Deriv{w_0^{\Ttop}}{r} & = \beta^{\Ttop} \\
   \Deriv{\beta^{\Ttop}}{r} & = - \cfrac{M_{rr}^{\Ttop}}{D_{11}^{\Ttop}} - \cfrac{D_{12}^{\Ttop}}{D_{11}^{\Ttop}}~\cfrac{\beta^{\Ttop}}{r} \\
  \Deriv{N^\Ttop_{rr}}{r} & = 
   \left[\cfrac{A^\Ttop_{12}-A^\Ttop_{11}}{A^\Ttop_{11}}\right]~\cfrac{N^\Ttop_{rr}}{r} +
   \left[\cfrac{(A^\Ttop_{11})^2 - (A^\Ttop_{12})^2}{A^\Ttop_{11}}\right]~\cfrac{u^\Ttop_{0r}}{r^2} 
   + \sigma^\Tcore_{rz} - s^\Ttt \\
  \Deriv{M^\Ttop_{rr}}{r} & = Q^\Ttop_r 
   + \left[\cfrac{D^\Ttop_{12}-D^{\Ttop}_{11}}{D^\Ttop_{11}}\right]~\cfrac{M^\Ttop_{rr}}{r} +
     \left[\cfrac{(D^\Ttop_{12})^2-(D^\Ttop_{11})^2}{D^\Ttop_{11}}\right]~\cfrac{\beta^\Ttop}{r^2} 
     + f^\Ttop~(s^\Ttt - \sigma^\Tcore_{rz}) \\
   \Deriv{Q_r^{\Ttop}}{r} & = -\cfrac{Q_r^\Ttop}{r}
     + \cfrac{C^{\Tcore}_{33}}{2c}~\left[w_0^\Ttop(r)-w_0^\Tbot(r)\right] 
     - c~T_r^\Tcore - c~\cfrac{\sigma^{\Tcore}_{rz}}{r} - q^\Ttt
     \end{aligned}
   \Eeq

   \item {\bf Bottom facesheet}:
   \Beq
     \begin{aligned}
   \Deriv{u^{\Tbot}_{0r}}{r} & = \cfrac{N_{rr}^{\Tbot}}{A_{11}^{\Tbot}} - \cfrac{A_{12}^{\Tbot}}{A_{11}^{\Tbot}}~\cfrac{u_{0r}^{\Tbot}}{r} \\
   \Deriv{w_0^{\Tbot}}{r} & = \beta^{\Tbot} \\
   \Deriv{\beta^{\Tbot}}{r} & = - \cfrac{M_{rr}^{\Tbot}}{D_{11}^{\Tbot}} - \cfrac{D_{12}^{\Tbot}}{D_{11}^{\Tbot}}~\cfrac{\beta^{\Tbot}}{r} \\
  \Deriv{N^\Tbot_{rr}}{r} & = 
   \left[\cfrac{A^\Tbot_{12}-A^\Tbot_{11}}{A^\Tbot_{11}}\right]~\cfrac{N^\Tbot_{rr}}{r} +
   \left[\cfrac{(A^\Tbot_{11})^2 - (A^\Tbot_{12})^2}{A^\Tbot_{11}}\right]~\cfrac{u^\Tbot_{0r}}{r^2}  
   - \sigma^\Tcore_{rz} - s^\Tbb\\
  \Deriv{M^\Tbot_{rr}}{r} & = Q^\Tbot_r
    + \left[\cfrac{D^\Tbot_{12}-D^{\Tbot}_{11}}{D^\Tbot_{11}}\right]~\cfrac{M^\Tbot_{rr}}{r} +
     \left[\cfrac{(D^\Tbot_{12})^2-(D^\Tbot_{11})^2}{D^\Tbot_{11}}\right]~\cfrac{\beta^\Tbot}{r^2} 
     - f^\Tbot~(s^\Tbb + \sigma^\Tcore_{rz}) \\
   \Deriv{Q_r^{\Tbot}}{r} & = -\cfrac{Q_r^\Tbot}{r} 
     - \cfrac{C^{\Tcore}_{33}}{2c}~\left[w_0^\Ttop(r)-w_0^\Tbot(r)\right] 
     - c~T_r^\Tcore - c~\cfrac{\sigma^{\Tcore}_{rz}}{r} - q^\Tbb
     \end{aligned}
   \Eeq

   \item {\bf Core}:
   \Beq
     \begin{aligned}
   \Deriv{\sigma^{\Tcore}_{rz}}{r} & = T_r^{\Tcore} \\
   \Deriv{T_r^{\Tcore}}{r} & =  
     -\cfrac{3C_{33}^{\Tcore}}{2c^3}~\left[u^{\Ttop}_{0r} - u_{0r}^{\Tbot} + 
        \left(f^{\Ttop}+c\right)~\beta^{\Ttop} + \left(f^{\Tbot} + c\right)~\beta^{\Tbot} \right]
        \\
     & \qquad 
     + \left(\cfrac{1}{r^2} + \cfrac{6C_{33}^{\Tcore}S_{55}^{\Tcore}}{c^2}\right)~\sigma^{\Tcore}_{rz} - \cfrac{T_r^{\Tcore}}{r}
     \end{aligned}
   \Eeq
\end{itemize}
\newglossaryentry{270}{name={$()^\Ttop$}, description={Related to the top facesheet}}
\newglossaryentry{271}{name={$()^\Tbot$}, description={Related to the bottom facesheet}}
\newglossaryentry{272}{name={$()^\Tcore$}, description={Related to the core}}

This is a set of 14 coupled ODEs that can be solved using a number of approaches.  Thomsen and
coworkers~\cite{Thomsen98,Thomsen98a} use a multi-segment integration approach to solve these
equations.  Since it is considerably simple to solve the original system of equations
using the finite element approach, we have used finite elements in this work.

\section{Finite element formulation of the coupled governing equations}
For the finite element formulation of the governing equations, it is convenient to start
with the statement of virtual work for the facesheets, i.e.,
\Beq
  \Bal
  \int_{\Omega_0} \left[N_{rr}~\Deriv{\delta u_{0r}}{r}  
    + \cfrac{N_{\theta\theta}}{r}~\delta u_{0r}  
    - M_{rr}~\DDeriv{\delta w_0}{r} 
    - \cfrac{M_{\theta\theta}}{r}~\Deriv{\delta w_0}{r} \right]~d\Omega_0
   = \\
     \qquad \int_{\Omega_0} \left[s(r)~\delta u_{0r} 
     - z_f~s(r)~\Deriv{\delta w_0}{r} + q(r)~\delta w_0\right]~d\Omega_0 \\
     + \oint_{\Gamma_0} \left[N_{r}~\delta u_{0r} - M_{r}~\Deriv{\delta w_0}{r}
        + Q_z~\delta w_0\right] ~d\Gamma_0
  \Eal
\Eeq
where
\Beq
  \Bal
  N_{rr} & = A_{11}~\Deriv{u_{0r}}{r} + A_{12}~\cfrac{u_{0r}}{r} ~;~~
  N_{\theta\theta}  = A_{12}~\Deriv{u_{0r}}{r} + A_{11}~\cfrac{u_{0r}}{r} \\
  M_{rr} & = -D_{11}~\DDeriv{w_{0}}{r} - \cfrac{D_{12}}{r}~\Deriv{w_{0}}{r} ~;~~
  M_{\theta\theta}  = -D_{12}~\DDeriv{w_{0}}{r} - \cfrac{D_{11}}{r}~\Deriv{w_{0}}{r} 
  \Eal
\Eeq
Separating terms containing $\delta u_{0r}$ and $\delta w_0$ leads to two equations
\begin{align}
  \int_{\Omega_0} \left[N_{rr}~\Deriv{\delta u_{0r}}{r}  
    + \left\{\cfrac{N_{\theta\theta}}{r} - s(r)\right\}~\delta u_{0r}\right]~d\Omega_0
   & = \oint_{\Gamma_0} N_{r}~\delta u_{0r} ~d\Gamma_0 \label{eq:Neq}\\
  \int_{\Omega_0} \left[
    M_{rr}~\DDeriv{\delta w_0}{r} 
    + \left\{\cfrac{M_{\theta\theta}}{r} + z_f~s(r)\right\}~\Deriv{\delta w_0}{r} 
    + q(r)~\delta w_0\right]~d\Omega_0
   & = \oint_{\Gamma_0} \left[ M_{r}~\Deriv{\delta w_0}{r}
        - Q_z~\delta w_0\right] ~d\Gamma_0 \label{eq:Meq}
\end{align}
The continuity of tractions across the facesheet-core interfaces requires that
\Beq
  \Bal
  q^\Ttop(r) = q^{\Ttb}(r) + q^{\Ttt}(r) & = -\cfrac{C^{\Tcore}_{33}}{2c}~[w^\Ttop_0(r)-w^\Tbot_0(r)] 
    +c~\Deriv{\sigma_{rz}^{\Tcore}}{r} +c~\cfrac{\sigma_{rz}^\Tcore}{r} + q^{\Ttt}(r) \\
  q^\Tbot(r) = q^{\Tbt}(r) + q^{\Tbb}(r) & = \cfrac{C^{\Tcore}_{33}}{2c}~[w^\Ttop_0(r)-w^\Tbot_0(r)] 
    +c~\Deriv{\sigma_{rz}^{\Tcore}}{r} +c~\cfrac{\sigma_{rz}^\Tcore}{r} + q^{\Tbb}(r)
  \Eal
\Eeq
and
\Beq
  \Bal
  s^\Ttop(r) = s^{\Ttb}(r) + s^{\Ttt}(r) & = -\sigma^{\Tcore}_{rz}(r) + s^{\Ttt}(r) \\
  s^\Tbot(r) = s^{\Tbt}(r) + s^{\Tbb}(r) & = \sigma^{\Tcore}_{rz}(r) + s^{\Tbb}(r) 
  \Eal
\Eeq
Plugging these into equations (\ref{eq:Neq}) and (\ref{eq:Meq}) leads to, for the top  
facesheet, 
\begin{align}
  \int_{\Omega_0} \biggl[N^{\Ttop}_{rr}~\Deriv{\delta u_{0r}^{\Ttop}}{r}  & 
    + \biggl\{\cfrac{N^{\Ttop}_{\theta\theta}}{r} + \sigma_{rz}^\Tcore - s^{\Ttt}\biggr\}
   ~\delta u^{\Ttop}_{0r}\biggr]~d\Omega_0
    = \oint_{\Gamma_0} N^{\Ttop}_{r}~\delta u^{\Ttop}_{0r} ~d\Gamma_0 \label{eq:NtopFS}\\
  \int_{\Omega_0} \biggl[ M^{\Ttop}_{rr}~\DDeriv{\delta w_0^{\Ttop}}{r}  & 
    + \biggl\{\cfrac{M^{\Ttop}_{\theta\theta}}{r} - f^\Ttop~\left(s^\Ttt - \sigma_{rz}^\Tcore\right) 
      \biggr\} ~\Deriv{\delta w_0^{\Ttop}}{r} \nonumber \\
    & - \biggl\{\cfrac{C^{\Tcore}_{33}}{2c}~(w^\Ttop_0-w^\Tbot_0) 
    -c~\Deriv{\sigma_{rz}^{\Tcore}}{r} -c~\cfrac{\sigma_{rz}^\Tcore}{r} 
      \biggr\}~\delta w^{\Ttop}_0\biggr]~d\Omega_0 \nonumber \\
    & \qquad \qquad = \oint_{\Gamma_0} \left[ M^{\Ttop}_{r}~\Deriv{\delta w_0^{\Ttop}}{r}
        - Q^{\Ttop}_z~\delta w_0^{\Ttop}\right] ~d\Gamma_0 \label{eq:MtopFS}
\end{align}
and for the bottom facesheet
\begin{align}
  \int_{\Omega_0} \biggl[N^{\Tbot}_{rr}~\Deriv{\delta u_{0r}^{\Tbot}}{r}  & 
    + \biggl\{\cfrac{N^{\Tbot}_{\theta\theta}}{r} - \sigma_{rz}^\Tcore - s^\Tbb \biggr\}
   ~\delta u^{\Tbot}_{0r}\biggr]~d\Omega_0
    = \oint_{\Gamma_0} N^{\Tbot}_{r}~\delta u^{\Tbot}_{0r} ~d\Gamma_0 \label{eq:NbotFS} \\
  \int_{\Omega_0} \biggl[ M^{\Tbot}_{rr}~\DDeriv{\delta w_0^{\Tbot}}{r}  & 
    + \biggl\{\cfrac{M^{\Tbot}_{\theta\theta}}{r} + f^\Tbot~\left(s^\Tbb + \sigma_{rz}^\Tcore\right)
      \biggr\} ~\Deriv{\delta w_0^{\Tbot}}{r} \nonumber \\
    & + \biggl\{\cfrac{C^{\Tcore}_{33}}{2c}~(w^\Ttop_0-w^\Tbot_0) 
    +c~\Deriv{\sigma_{rz}^{\Tcore}}{r} +c~\cfrac{\sigma_{rz}^\Tcore}{r} 
      \biggr\}~\delta w^{\Tbot}_0\biggr]~d\Omega_0 \nonumber \\
    & \qquad \qquad = \oint_{\Gamma_0} \left[ M^{\Tbot}_{r}~\Deriv{\delta w_0^{\Tbot}}{r}
        - Q^{\Tbot}_z~\delta w_0^{\Tbot}\right] ~d\Gamma_0 \label{eq:MbotFS}  
\end{align}

The governing ordinary differential equation for the core is
\Beq 
   \Bal
   \DDeriv{\sigma_{rz}^\Tcore}{r} & + \cfrac{1}{r}~\Deriv{\sigma_{rz}^\Tcore}{r}
     - \left(\cfrac{1}{r^2} + \cfrac{6C_{33}^\Tcore S_{55}^\Tcore}{c^2}\right)~\sigma_{rz}^\Tcore \\
     & \qquad = -\cfrac{3C_{33}^\Tcore}{2c^3}~\left[u^{\Ttop}_{0r} - u_{0r}^{\Tbot} 
       + \left(f^{\Ttop}+c\right)~\Deriv{w_0^{\Ttop}}{r} 
       + \left(f^{\Tbot} + c\right)~\Deriv{w_0^{\Tbot}}{r} \right]
   \Eal
\Eeq
Multiplying the equation with a test function and integration over the area $\Omega_0$ yields,
after an integration by parts, the equation:
\Beq \label{eq:SigCore}
 \Bal
  \int_{\Omega_0} \biggl[ \Deriv{\sigma_{rz}^\Tcore}{r}~\Deriv{\delta \sigma_{rz}}{r} & + 
    \cfrac{1}{r} \left( \sigma_{rz}^\Tcore~\Deriv{\delta \sigma_{rz}}{r} +
    \Deriv{\sigma_{rz}^\Tcore}{r}~\delta \sigma_{rz} \right) + 
   \biggl\{\left(\cfrac{1}{r^2} +
    \cfrac{6C_{33}^\Tcore S_{55}^\Tcore}{c^2}\right)~\sigma_{rz}^\Tcore \\
    & - \cfrac{3C_{33}^\Tcore}{2c^3}~\left[u^{\Ttop}_{0r} -
      u_{0r}^{\Tbot} +  \left(f^{\Ttop}+c\right)~\Deriv{w_0^{\Ttop}}{r} +
      \left(f^{\Tbot} + c\right)~\Deriv{w_0^{\Tbot}}{r} \right]\biggr\}~\delta \sigma_{rz}
    \biggr]~d\Omega_0 \\
   & = \oint_{\Gamma_0} 
    \left(\Deriv{\sigma_{rz}^\Tcore}{r} + \cfrac{\sigma_{rz}^\Tcore}{r} \right)~
     \delta\sigma_{rz}~d\Gamma_0  
  \Eal
\Eeq

Equations (\ref{eq:NtopFS}), (\ref{eq:MtopFS}), (\ref{eq:NbotFS}), (\ref{eq:MbotFS}), and
(\ref{eq:SigCore}) form the system that has been discretized using the finite element approach.

We assume that the fields $u_{0r}^\Ttop, u_{0r}^\Tbot, w_0^\Ttop, w_0^\Tbot, \sigma_{rz}^\Tcore$
can be expressed as
\Beq
  \Bal
     u_{0r}^\Ttop(r) & = \sum_{i=1}^{nu} u_i^\Ttop~N^u_i(r) ~;~~
     u_{0r}^\Tbot(r) = \sum_{i=1}^{nu} u_i^\Tbot~N^u_i(r) \\
     w_{0}^\Ttop(r) & = \sum_{i=1}^{nw} w_i^\Ttop~N^w_i(r) ~;~~
     w_{0}^\Tbot(r) = \sum_{i=1}^{nw} w_i^\Tbot~N^w_i(r) \\
     \sigma_{rz}^\Tcore(r) & = \sum_{i=1}^{ns} \sigma_i~N^s_i(r)
  \Eal
\Eeq
where $nu, nw, ns$ are the number of nodes and $N^{u,w,s}_i$ are the basis functions that are 
required to represent the field variables.  Then, the stress and stress couple resultants
can be expressed as
\Beq
  \Bal
  N^\Ttop_{rr} & = \sum_{j=1}^{nu} \left[A^\Ttop_{11}~\Deriv{N^u_j}{r} + 
                                         A^\Ttop_{12}~\cfrac{N^u_j}{r}\right]~u_j^\Ttop ~;~~
  N^\Ttop_{\theta\theta} = \sum_{j=1}^{nu} \left[A^\Ttop_{12}~\Deriv{N^u_j}{r} + 
                                         A^\Ttop_{11}~\cfrac{N^u_j}{r}\right]~u_j^\Ttop \\
  N^\Tbot_{rr} & = \sum_{j=1}^{nu} \left[A^\Tbot_{11}~\Deriv{N^u_j}{r} + 
                                         A^\Tbot_{12}~\cfrac{N^u_j}{r}\right]~u_j^\Tbot ~;~~
  N^\Tbot_{\theta\theta} = \sum_{j=1}^{nu} \left[A^\Tbot_{12}~\Deriv{N^u_j}{r} + 
                                         A^\Tbot_{11}~\cfrac{N^u_j}{r}\right]~u_j^\Tbot \\
  M^\Ttop_{rr} & = -\sum_{j=1}^{nw} \left[D^\Ttop_{11}~\DDeriv{N^w_j}{r} + 
                                          \cfrac{D^\Ttop_{12}}{r}~\Deriv{N^w_j}{r}\right]
                                          ~w_j^\Ttop ~;~~
  M^\Ttop_{\theta\theta}  = -\sum_{j=1}^{nw} \left[D^\Ttop_{12}~\DDeriv{N^w_j}{r} + 
                                          \cfrac{D^\Ttop_{11}}{r}~\Deriv{N^w_j}{r}\right]
                                          ~w_j^\Ttop \\
  M^\Tbot_{rr} & = -\sum_{j=1}^{nw} \left[D^\Tbot_{11}~\DDeriv{N^w_j}{r} + 
                                          \cfrac{D^\Tbot_{12}}{r}~\Deriv{N^w_j}{r}\right]
                                          ~w_j^\Tbot ~;~~
  M^\Tbot_{\theta\theta}  = -\sum_{j=1}^{nw} \left[D^\Tbot_{12}~\DDeriv{N^w_j}{r} + 
                                          \cfrac{D^\Tbot_{11}}{r}~\Deriv{N^w_j}{r}\right]
                                          ~w_j^\Tbot 
  \Eal
\Eeq
and the momentum balance equations can be written as
\begin{align}
  &\int_{\Omega_0} \biggl[N^{\Ttop}_{rr}~\Deriv{N^u_i}{r}  
    + \biggl\{\cfrac{N^{\Ttop}_{\theta\theta}}{r} + \sum_{k=1}^{ns} \sigma_k~N_k^s\biggr\}
   ~N_i^u\biggr]~d\Omega_0
     = \int_{\Omega_0} s^\Ttt~N_i^u~d\Omega_0 
       + \oint_{\Gamma_0} N^{\Ttop}_{r}~N_i^u~d\Gamma_0 \\
  &\int_{\Omega_0} \biggl[N^{\Tbot}_{rr}~\Deriv{N^u_i}{r}  
    + \biggl\{\cfrac{N^{\Tbot}_{\theta\theta}}{r} - \sum_{k=1}^{ns} \sigma_k~N_k^s\biggr\}
   ~N_i^u\biggr]~d\Omega_0
     = \int_{\Omega_0} s^\Tbb~N_i^u~d\Omega_0 
       + \oint_{\Gamma_0} N^{\Tbot}_{r}~N_i^u~d\Gamma_0 \\
  &\int_{\Omega_0} \biggl[M^{\Ttop}_{rr}~\DDeriv{N_i^w}{r}  
    + \biggl\{\cfrac{M^{\Ttop}_{\theta\theta}}{r} + f^\Ttop~\sum_{k=1}^{ns}\sigma_k~N_k^s 
      \biggr\}~\Deriv{N_i^w}{r}  
     - \biggl\{\cfrac{C^{\Tcore}_{33}}{2c}~\sum_{k=1}^{nu} (w_k^\Ttop-w_k^\Tbot)~N_k^w 
     \nonumber \\
  &   \qquad \qquad \qquad -c~\sum_{k=1}^{ns}\left(\Deriv{N_k^s}{r} + \cfrac{N_k^s}{r}\right)\sigma_k
     \biggr\}~N_i^w\biggr]~d\Omega_0 \nonumber \\
  & \qquad    = \int_{\Omega_0} f^\Ttop~s^\Ttt~\Deriv{N_i^w}{r}~d\Omega_0 
      + \oint_{\Gamma_0} \left[ M^{\Ttop}_{r}~\Deriv{N_i^w}{r}
        - Q^{\Ttop}_z~N_i^w\right] ~d\Gamma_0 \\
  &\int_{\Omega_0} \biggl[M^{\Tbot}_{rr}~\DDeriv{N_i^w}{r}  
    + \biggl\{\cfrac{M^{\Tbot}_{\theta\theta}}{r} + f^\Tbot~\sum_{k=1}^{ns}\sigma_k~N_k^s 
      \biggr\}~\Deriv{N_i^w}{r}  
     + \biggl\{\cfrac{C^{\Tcore}_{33}}{2c}~\sum_{k=1}^{nu} (w_k^\Ttop-w_k^\Tbot)~N_k^w 
     \nonumber \\
  &   \qquad \qquad \qquad +c~\sum_{k=1}^{ns}\left(\Deriv{N_k^s}{r} + \cfrac{N_k^s}{r}\right)\sigma_k
     \biggr\}~N_i^w\biggr]~d\Omega_0 \nonumber \\
  & \qquad    = - \int_{\Omega_0} f^\Tbot~s^\Tbb~\Deriv{N_i^w}{r}~d\Omega_0 
      + \oint_{\Gamma_0} \left[ M^{\Tbot}_{r}~\Deriv{N_i^w}{r}
        - Q^{\Tbot}_z~N_i^w\right] ~d\Gamma_0 \\
  &\int_{\Omega_0} \biggl[\sum_{j=1}^{ns}\biggl\{\left(\Deriv{N_j^s}{r}+\cfrac{N_j^s}{r}\right)
     ~\Deriv{N_i^s}{r} + \Deriv{N_j^s}{r}~\cfrac{N_i^s}{r}  + 
      \biggl(\cfrac{1}{r^2} + \cfrac{6C_{33}^\Tcore S_{55}^\Tcore}{c^2}\biggr)~N_j^s~N_i^s\biggr\}
      \sigma_j \nonumber \\  
  &   \qquad - \cfrac{3C_{33}^\Tcore}{2c^3}~\left[\sum_{k=1}^{nu} N_k^u~(u_k^{\Ttop} - u_k^{\Tbot})  
    +  \sum_{k=1}^{nw} \Deriv{N_k^w}{r}
        \biggl\{\left(f^{\Ttop}+c\right)~w_k^{\Ttop} +
                \left(f^{\Tbot}+c\right)~w_k^{\Tbot}\biggr\} \right]~N_i^s
    \biggr]~d\Omega_0  \nonumber \\
  & \qquad \qquad   = \oint_{\Gamma_0} 
    \left(\Deriv{\sigma_{rz}^\Tcore}{r} + \cfrac{\sigma_{rz}^\Tcore}{r} \right)~
     N_i^s~d\Gamma_0  
\end{align}
After plugging in the expressions for the resultant stress and stress couples, we can
express the above equations in matrix form as
\Beq
  \begin{bmatrix} 
    [\mathbf{K}^{tt}_{uu}] & [\mathbf{0}] & [\mathbf{0}] & 
       [\mathbf{0}] & [\mathbf{K}^{t}_{us}] \\
    [\mathbf{0}] & [\mathbf{K}^{tt}_{ww}] & [\mathbf{0}] & 
       [\mathbf{0}] & [\mathbf{K}^{t}_{ws}] \\
    [\mathbf{0}] & [\mathbf{0}] & [\mathbf{K}^{bb}_{uu}] & 
       [\mathbf{0}] & [\mathbf{K}^{b}_{us}] \\
    [\mathbf{0}] & [\mathbf{0}] & [\mathbf{0}] & 
       [\mathbf{K}^{bb}_{ww}] & [\mathbf{K}^{b}_{ws}] \\
    [\mathbf{K}^{t}_{su}] & [\mathbf{K}^{t}_{sw}] & [\mathbf{K}^{b}_{su}] & 
       [\mathbf{K}^{b}_{sw}] & [\mathbf{K}_{ss}] 
  \end{bmatrix}
  \begin{bmatrix}
    \mathbf{u}^\Ttop \\ \mathbf{w}^\Ttop \\ \mathbf{u}^\Tbot \\ \mathbf{w}^\Tbot \\ 
      \boldsymbol{\sigma}
  \end{bmatrix}
  = 
  \begin{bmatrix}
    \mathbf{f}^\Ttop_u \\ \mathbf{f}^\Ttop_w \\ 
    \mathbf{f}^\Tbot_u \\ \mathbf{f}^\Tbot_w \\ \mathbf{f}_\sigma
  \end{bmatrix}
\Eeq

\subsection{Finite element basis functions}
Note that the stiffness matrix is not symmetric.  This system of equations is solved
using COMSOL\texttrademark using quadratic shape functions for the $u$-displacement
and the $\sigma$-stress and cubic Hermite functions for the $w$-displacement, i.e.,
in each element $nu = ns = 3$, $nw = 4$, and
\Beq
   \Bal
   N_i^u(r) & = N_i^s(r) = \prod_{j=1, i\ne j}^{3} \cfrac{r - r_j}{r_i - r_j} \\
   N_1^w(r) & = 1 - 3~r^2 + 2~r^3 ~;~~ 
   N_2^w(r) = (r_2-r_1)\left(r - 2~r^2 + r^3\right) \\
   N_3^w(r) & = 3~r^2 - 2~r^3 ~;~~
   N_4^w(r) = (r_2-r_1)(-r^2 + r^3)
   \Eal
\Eeq

\subsection{Boundary conditions}
The natural boundary conditions are
\Beq
  \Bal
  & \mbox{Top~facesheet}    :~~ N_r^\Ttop, Q_z^\Ttop, M_r^\Ttop  \\
  & \mbox{Bottom~facesheet} :~~ N_r^\Tbot, Q_z^\Tbot, M_r^\Tbot  \\
  & \mbox{Core}             :~~ \Deriv{\sigma_{rz}^\Tcore}{r} + \cfrac{\sigma_{rz}^\Tcore}{r} 
  \Eal
\Eeq
The essential boundary conditions are
\Beq
  \Bal
  & \mbox{Top~facesheet}    :~~ u_{0r}^\Ttop, w_0^\Ttop, \Deriv{w_0^\Ttop}{r} \\
  & \mbox{Bottom~facesheet} :~~ u_{0r}^\Tbot, w_0^\Tbot, \Deriv{w_0^\Tbot}{r} \\
  & \mbox{Core}             :~~ \sigma_{rz}^\Tcore
  \Eal
\Eeq
Note that fixing the $u_z$ displacement at the boundary of the core is equivalent to
setting the natural boundary condition in the core to zero when $w_0^\Ttop = w_0^\Tbot=0$
at the boundary.

\subsubsection{Through-the-thickness insert}
\begin{figure}[htb!]
  \centering
  \scalebox{0.4}{\includegraphics{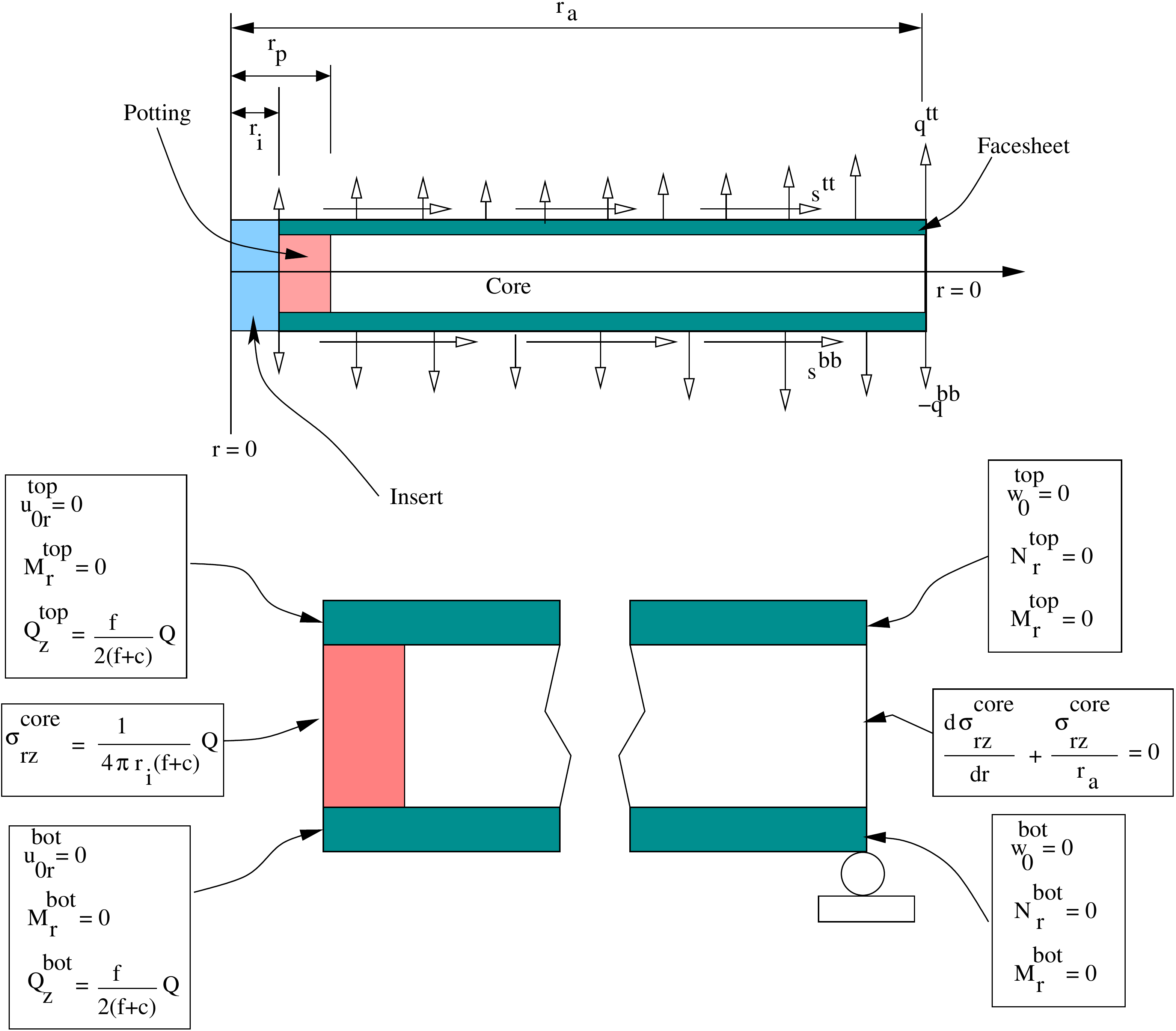}}
  \caption{Boundary conditions for a through-the-thickness insert in a
           simply supported sandwich panel.}
  \label{fig:ThruInsertBCs}
\end{figure}

The boundary conditions used for a simply-supported sandwich panel with a
through-the-thickness insert are shown in Figure~\ref{fig:ThruInsertBCs}.
The radius of the insert is $r_i$, that of the potting is $r_p$, and that 
of the panel is $r_a$.  Therefore, for part of the panel, the potting is
assumed to have the same behavior as the core.

For this situation, the boundary conditions at the left edge, $r = r_i$, are
\Beq
  \Bal
  u_{0r}^{\Ttop} & = u_{0r}^{\Tbot} = 0 ~;~~ M_r^{\Ttop} = M_r^{\Tbot} = 0 \\
  Q_z^{\Ttop} & = \cfrac{Q~f^{\Ttop}}{f^{\Ttop}+2c+f^{\Tbot}} ~;~~ 
  Q_z^{\Tbot} = \cfrac{Q~f^{\Tbot}}{f^{\Ttop}+2c+f^{\Tbot}} \\
  \sigma^\Tcore_{rz} & = \cfrac{Q}{A},
  \Eal
\Eeq
where $A=2~\pi~r_i~(f^{\Ttop}+2c+f^{\Tbot})$.  These are applied to the two
dimensional model as a constant pressure in the $z$ direction, with $P=Q/A$.

At the right edge, $r=r_a$, the structure is simply supported, with the
conditions: 
\Beq
  w^{\Ttop} = w^{\Tbot} = 0 ~;~~
  M_r^{\Ttop} = M_r^{\Tbot} = 0 ~;~~
  N_r^{\Ttop} = N_r^{\Tbot} = 0 ~;~~
 \Deriv{\sigma_{rz}}{r} + \cfrac{\sigma_{rz}}{r_a} = 0.
\Eeq
The support condition is applied to the two dimensional model by setting $w_0=0$
along the right edge.

\subsubsection{Potted insert}
\begin{figure}[htb!]
  \centering
  \scalebox{0.4}{\includegraphics{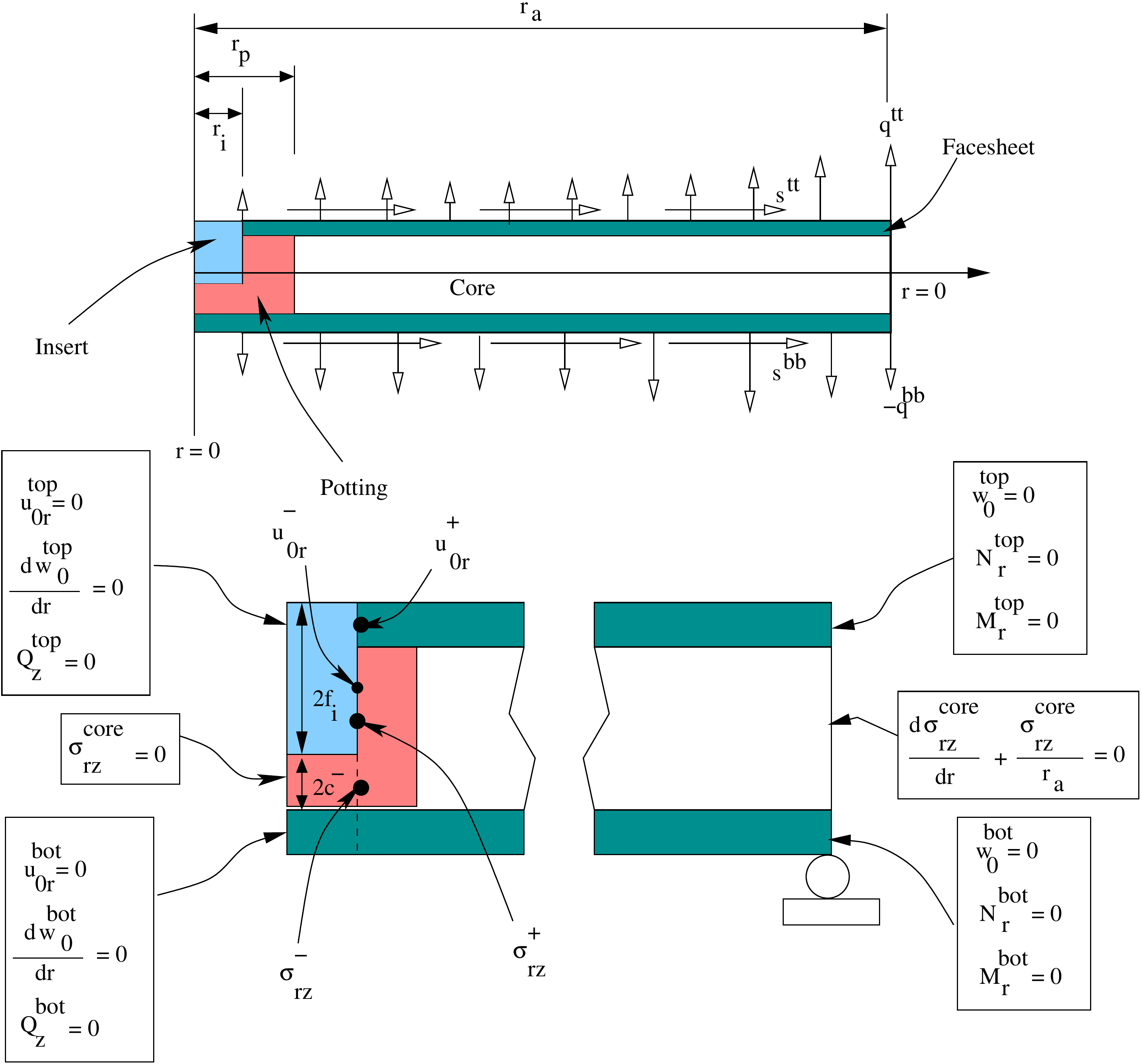}}
  \caption{Boundary conditions for a potted insert in a
           simply supported sandwich panel.}
  \label{fig:PartInsertBCs}
\end{figure}

The boundary conditions used for a simply-supported sandwich panel with a
potted insert are shown in Figure~\ref{fig:PartInsertBCs}.
The radius of the insert is $r_i$, that of the potting is $r_p$, and that 
of the panel is $r_a$.  The length of the insert is $2f_i$ and the thickness of 
the potting below the insert is $2c^{-}$.  Therefore, the insert is being 
treated as a thin plate in the region above the potting and the potting is
being treated as a material with features similar to the core.

To allow for the jump discontinuities on the two sides of the insert-facesheet
interface, we define the quantities $u_{0r}^{-}$ and $u_{0r}^{+}$ to be the 
$u_{0r}$ displacements of the insert and the top facesheet, respectively.
The locations where these quantities are evaluated are shown in 
Figure~\ref{fig:PartInsertBCs}.  Then the continuity of displacements
requires that
\Beq
  u_{0r}^{+} = u_{0r}^{-} - (f_i - f)~\Deriv{w_0^{\text{interface}}}{r}
\Eeq

There is also a jump in the shear stress in the two sections of the 
potting to the left and right of the interface.  Let these quantities 
be $\sigma_{rz}^{-}$ and $\sigma_{rz}^{+}$.  We assume that the average
force at the interface is balanced, i.e.,
\Beq
  c^{-}~\sigma_{rz}^{-} = c~\sigma_{rz}^{+}~.
\Eeq
The boundary conditions at $r = 0$ are
\Beq
  u_{0r}^{\Ttop}  = u_{0r}^{\Tbot} = 0 ~;~~ 
  \Deriv{w_0^\Ttop}{r} = \Deriv{w_0^\Tbot}{r} = 0 ~;~~
  Q_z^{\Ttop} = Q_z^{\Tbot} = 0  ~;~~
  \sigma^\Tcore_{rz}  = 0
\Eeq
The simply-supported boundary at $r = r_a$ once again requires that
\Beq
  w^{\Ttop} = w^{\Tbot} = 0 ~;~~
  M_r^{\Ttop} = M_r^{\Tbot} = 0 ~;~~
  N_r^{\Ttop} = N_r^{\Tbot} = 0 ~;~~
 \Deriv{\sigma_{rz}}{r} + \cfrac{\sigma_{rz}}{r_a} = 0.
\Eeq

\section{Model Test Cases: FRP Sandwich}
In order to validate the one dimensional approximation, the results for test
cases are compared with the results generated by a two dimensional
axisymmetric model.  In each test case, a rigid, through the thickness insert
applies a vertical compression load of $Q = 1000\text{N}$ to a simply supported sandwich
panel.

The facesheets are assumed to be isotropic, i.e,
\Beq
  \Bal
  A_{11} = 2~f~C_{11} = 2f~\cfrac{E}{1-\nu^2} ~;~~
  A_{12} = 2~f~C_{12} = 2f~\cfrac{\nu~E}{1-\nu^2} \\
  D_{11} = \cfrac{2~f^3}{3}~C_{11} = \cfrac{2f^3}{3}~\cfrac{E}{1-\nu^2} ~;~~
  D_{12} = \cfrac{2~f^3}{3}~C_{12} = \cfrac{2f^3}{3}~\cfrac{\nu~E}{1-\nu^2} 
  \Eal
\Eeq
where $E$ is the Young's modulus and $\nu$ is the Poisson's ratio.

The core moduli are given by
\Beq
  C_{33} = \cfrac{E_h}{1-\nu_h^2} ~;~~ C_{55} = 2~G_h
\Eeq
where $E_h, G_h, \nu_h$ are the Young's modulus, shear modulus, and the Poisson's ratio
of the core.  The potting moduli are also computed in a manner similar to those of the 
core.

\subsection{Example 1: Stiff facesheets}
\begin{figure}[htb!]
  \begin{minipage}[b]{0.5\linewidth}
    \centering
    \scalebox{0.45}{\includegraphics{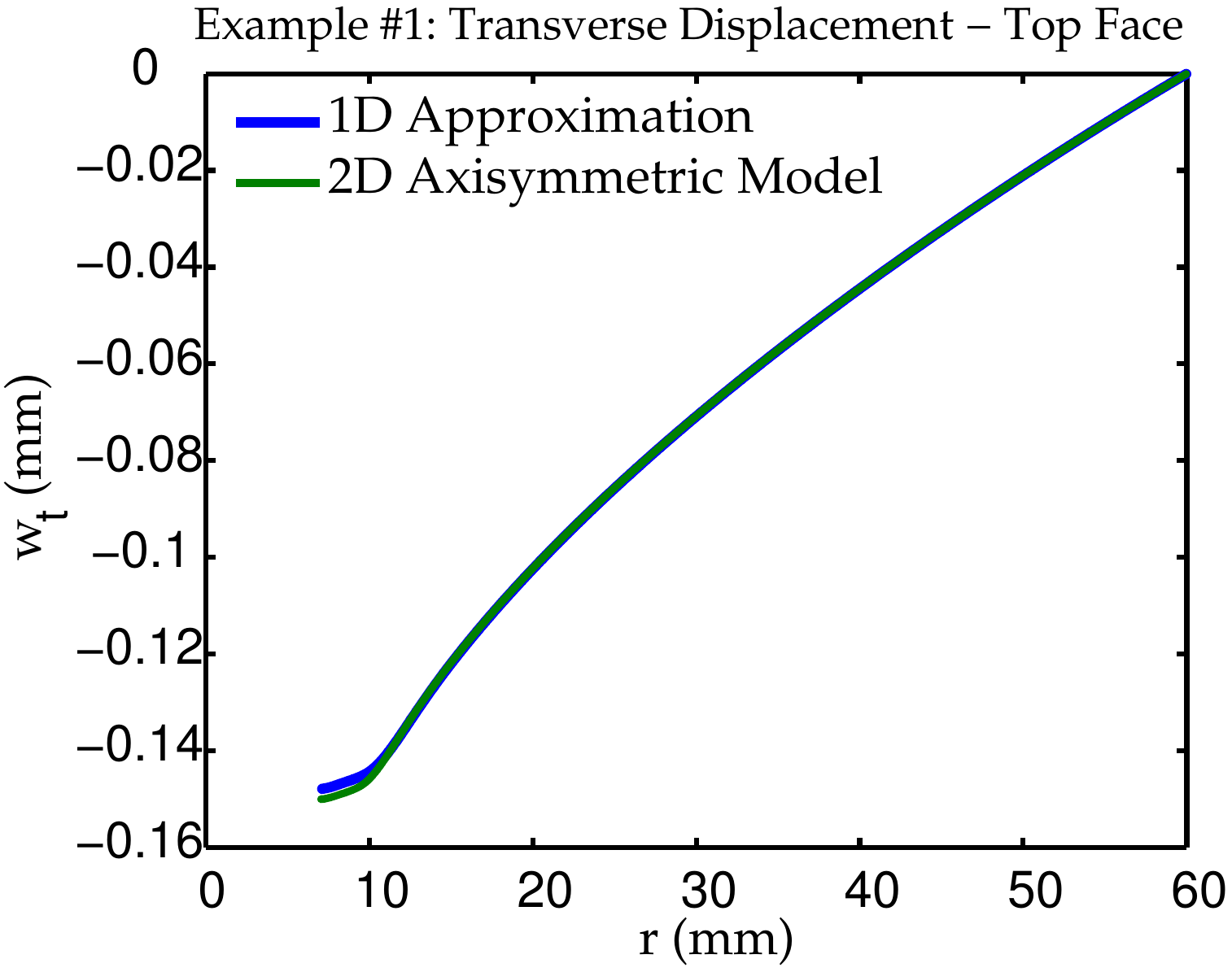}}\\
    (a) Out-of-plane displacement ($w_t$).
  \end{minipage}
  \begin{minipage}[b]{0.5\linewidth}
    \centering
    \scalebox{0.45}{\includegraphics{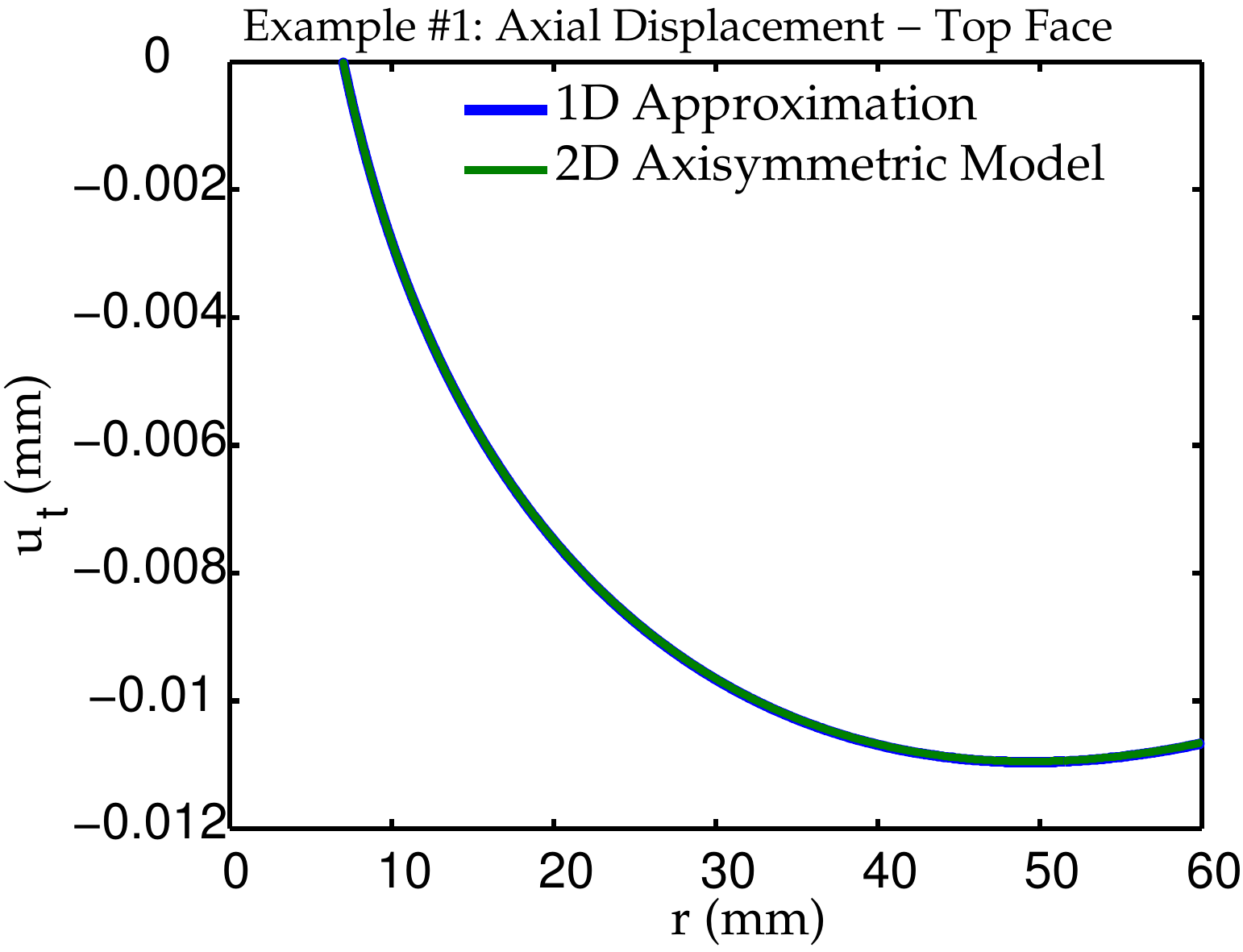}}\\
    (b) In-plane displacement ($u_t$).
  \end{minipage}
  \caption{Comparisons of displacements from one-dimensional and two-dimensional
    finite element simulations for the model in Table~\ref{tab:Ex1}.}
  \label{fig:w-Ex1}
\end{figure}

\begin{figure}[htb!]
  \begin{minipage}[b]{0.5\linewidth}
    \centering
    \scalebox{0.45}{\includegraphics{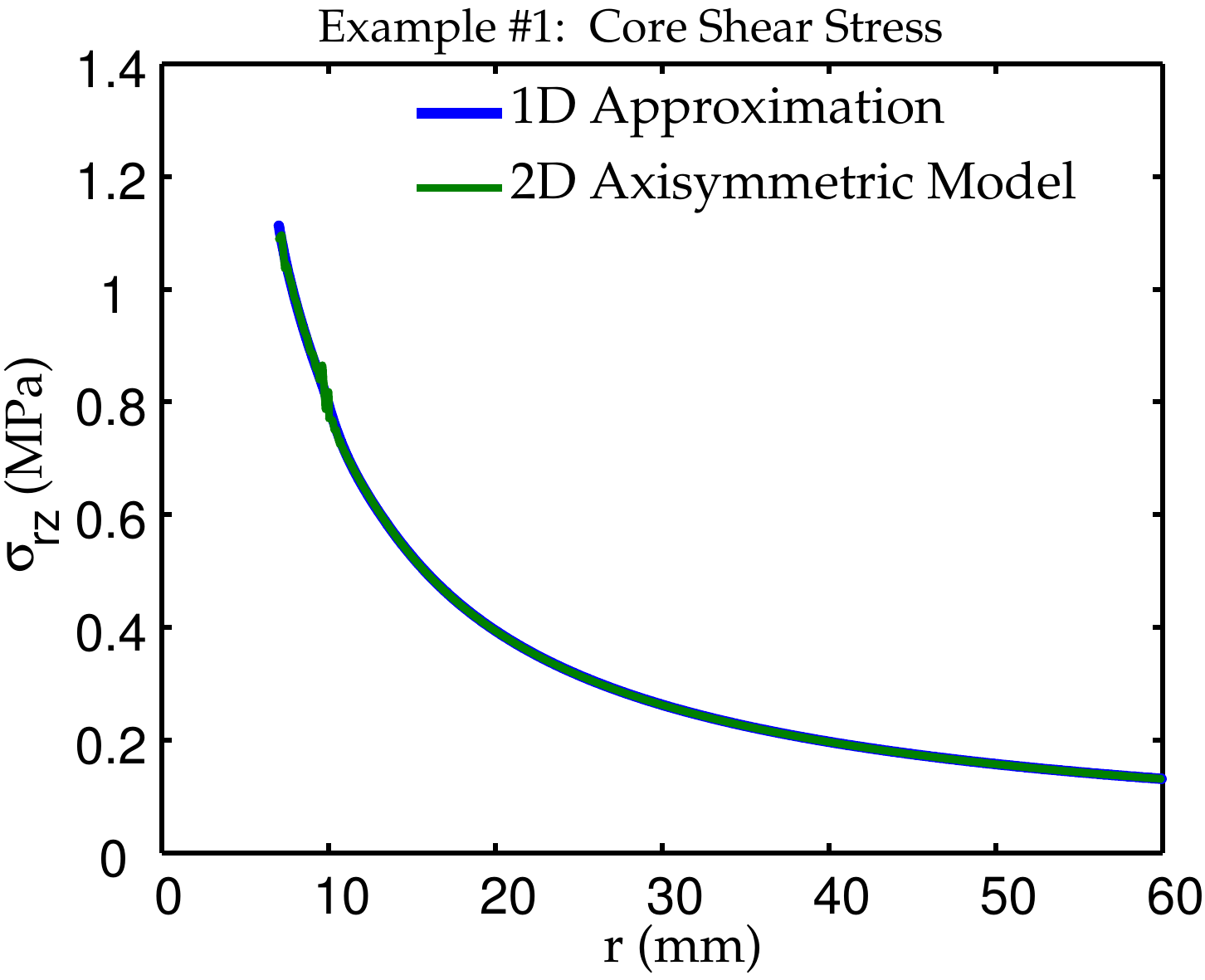}}\\
    (a) Core shear stress ($\sigma_{rz}$).\hspace{48pt}
  \end{minipage}
  \begin{minipage}[b]{0.5\linewidth}
    \centering
    \scalebox{0.45}{\includegraphics{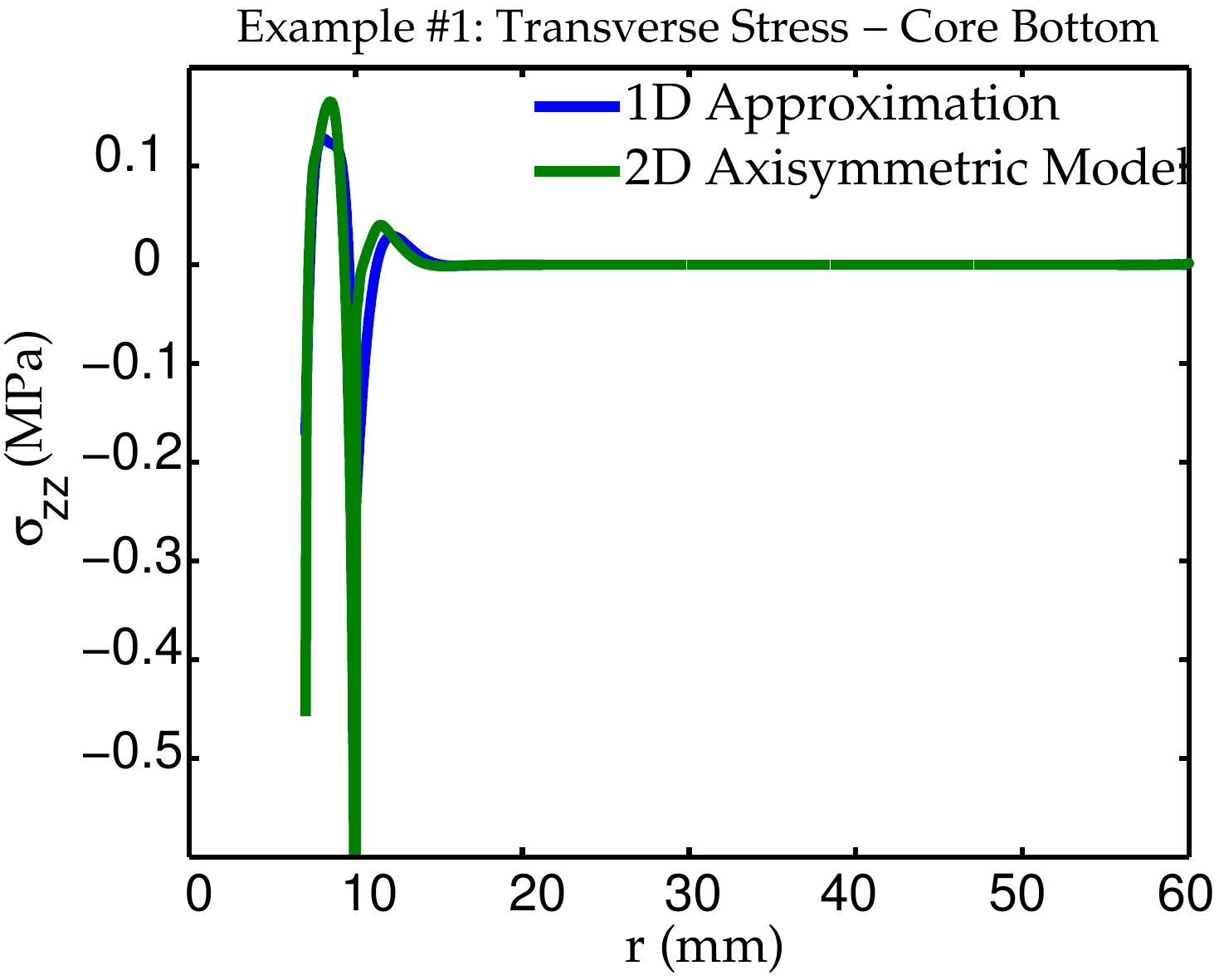}}\\
    (b) Transverse stress ($\sigma_{zz}$)- core bottom.
  \end{minipage}
  \caption{Comparisons of stresses from one-dimensional and two-dimensional
    finite element simulations for the model in Table~\ref{tab:Ex1}.}
  \label{fig:srz-Ex1}
\end{figure}

The first example problem is taken from \cite{Thomsen98}, with the parameters
given in Table \ref{tab:Ex1}.
Figure \ref{fig:w-Ex1}(a) compares the resulting out of plane displacements from
the sandwich theory and the two dimensional axisymmetric simulations.  While
there is a small amount of disagreement in the potting region, the overall
results match up well.  The radial displacement, $u_r$, is shown in figure
\ref{fig:w-Ex1}(b), and these results match as well.  Core shear stresses
and transverse stresses at the bottom of the core are shown in 
Figure~\ref{fig:srz-Ex1}.  The stresses match reasonably well too.
\begin{table}[ht!]
  \centering
  \caption{Geometric and material parameters for Example 1} \label{tab:Ex1}
  \begin{tabular}{lllllll}
    Geometry (mm): & $r_i=7.0$ & $r_p=10.0$ & $r_a=60.0$ & $c=10.0$ 
      & $f^{\Ttop}=0.1$ & $f^{\Tbot}=0.1$ \\
    Face Sheets (GPa) : & $E_1=71.5$ & $G_1=27.5$ \\
    Potting (GPa) :  &  $E_p=2.5$ & $G_p=0.93$ \\
    Honeycomb (MPa) : & $E_h=310$ & $G_h=138$
  \end{tabular}
\end{table}

  \subsection{Example 2: Soft facesheets}
\begin{figure}[htbp!]
    \begin{minipage}[b]{0.5\linewidth}
      \centering
      \scalebox{0.45}{\includegraphics{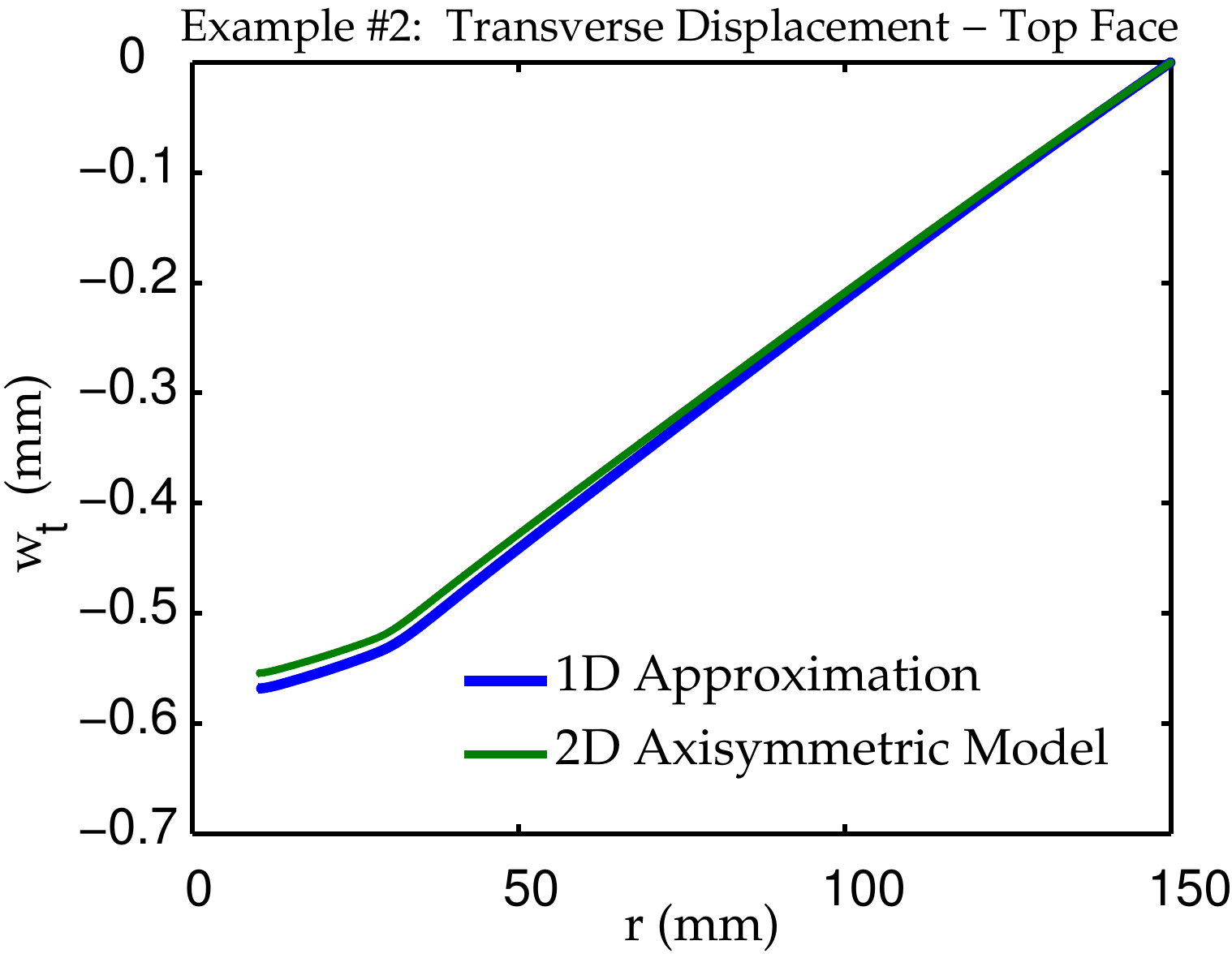}}\\
      (a) Out-of-plane displacement ($w_t$).
    \end{minipage}
    \begin{minipage}[b]{0.5\linewidth}
      \centering
      \scalebox{0.45}{\includegraphics{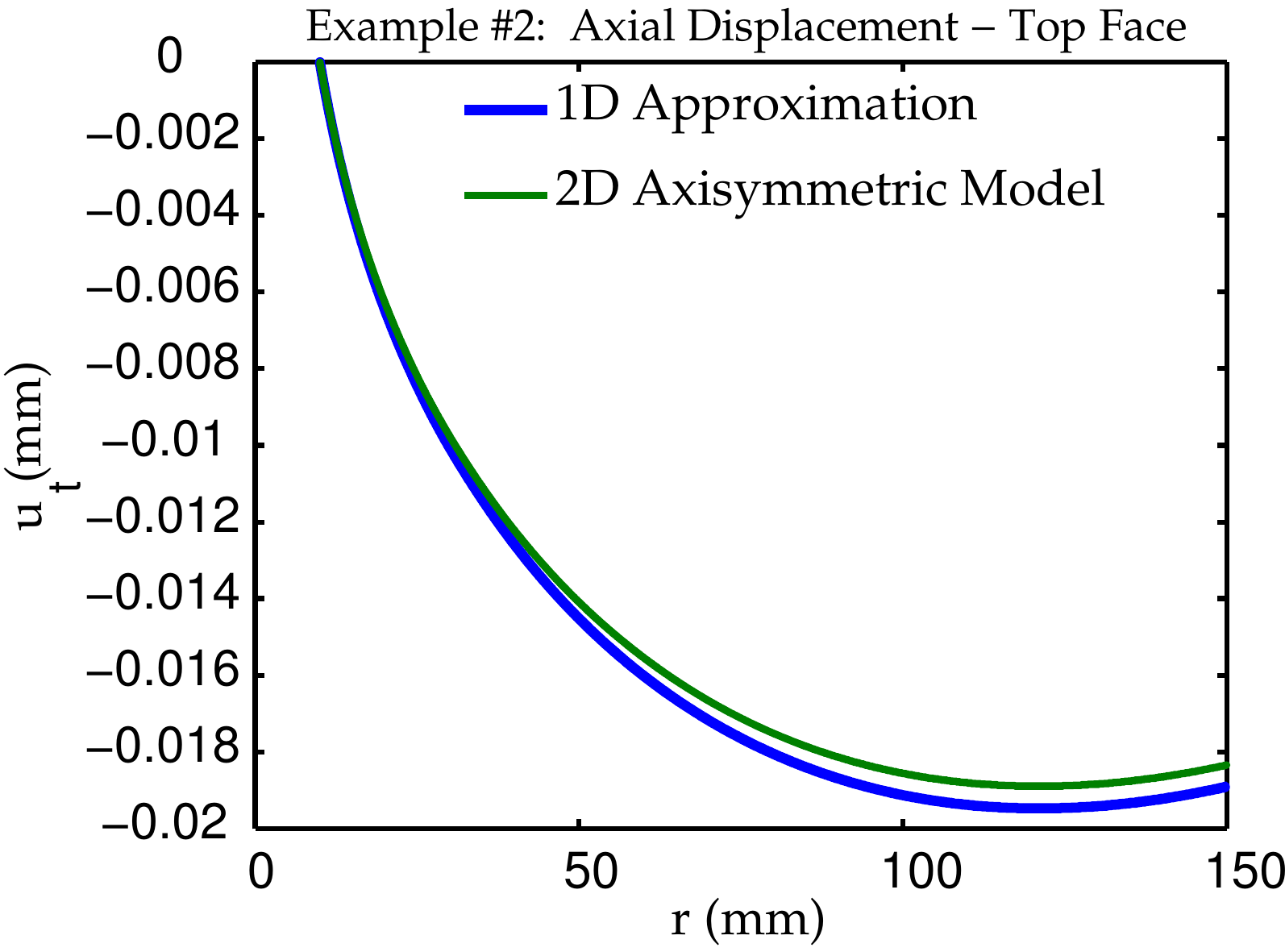}}\\
      (b) In-plane displacement ($u_t$).
    \end{minipage}
    \caption{Comparison of displacements from one-dimensional and two-dimensional
      finite element simulations for the model in Table~\ref{tab:Ex2}.}
    \label{fig:w-Ex2}
  \end{figure}

\begin{figure}[htbp!]
    \begin{minipage}[b]{0.5\linewidth}
      \centering
      \scalebox{0.45}{\includegraphics{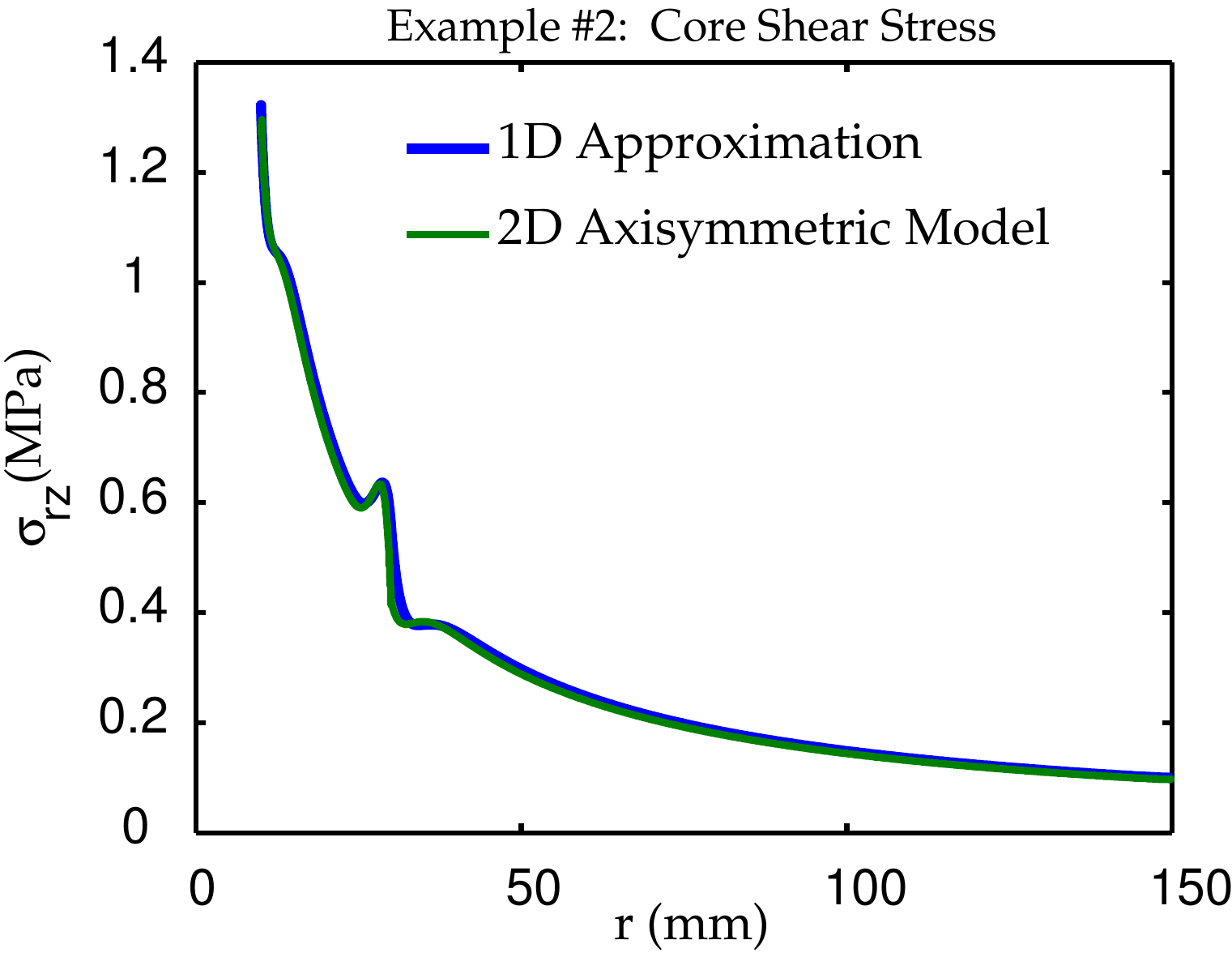}}\\
      (a) Core shear stress ($\sigma_{rz}$).
    \end{minipage}
    \begin{minipage}[b]{0.5\linewidth}
      \centering
      \scalebox{0.45}{\includegraphics{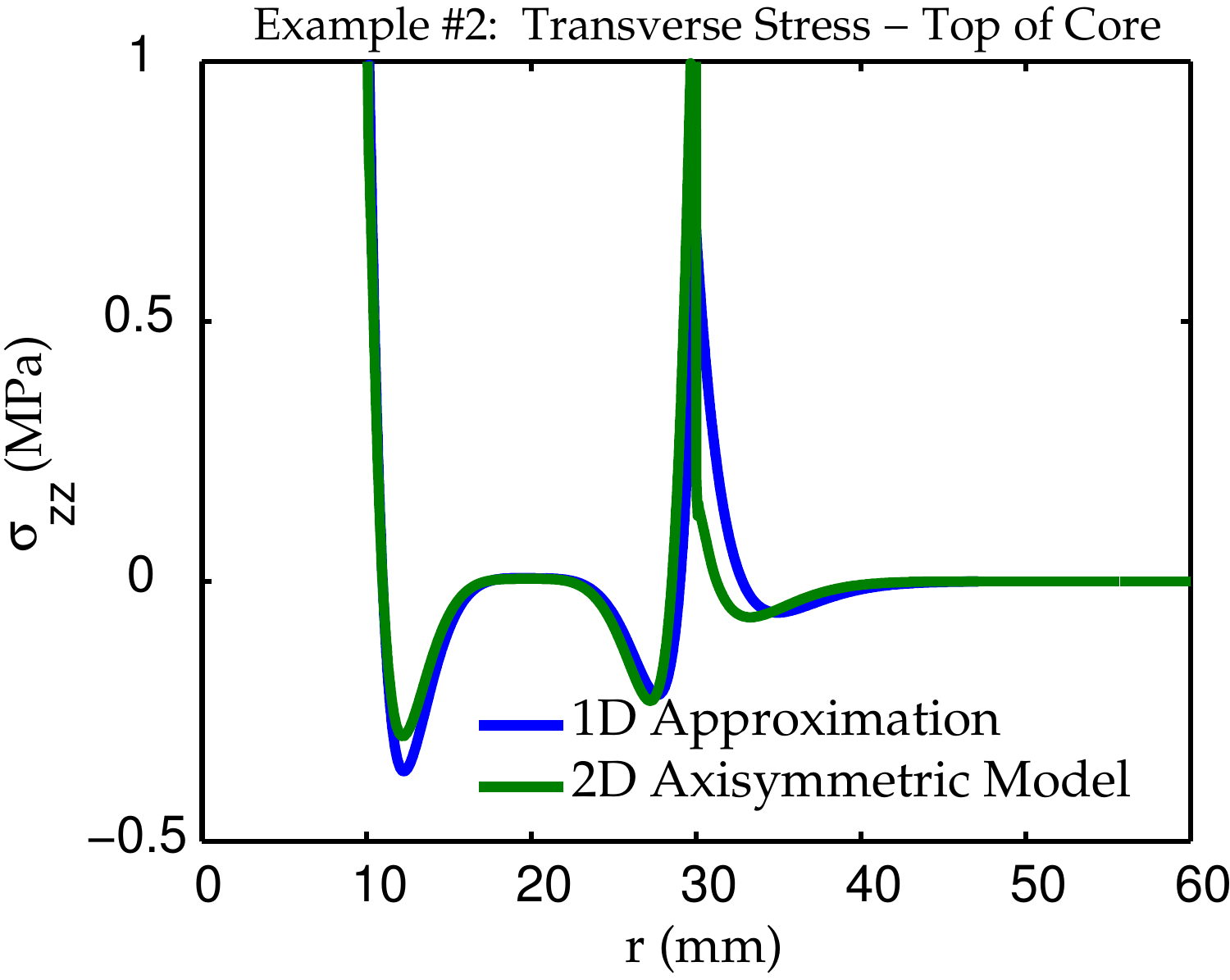}}\\
      (b) Transverse stress ($\sigma_{zz}$)- core top.
    \end{minipage}
    \caption{Comparison of stresses from one-dimensional and two-dimensional
      finite element simulations for the model in Table~\ref{tab:Ex2}.}
    \label{fig:srz-Ex2}
  \end{figure}

\begin{figure}[htbp!]
    \begin{minipage}[b]{0.5\linewidth}
      \centering
      \scalebox{0.45}{\includegraphics{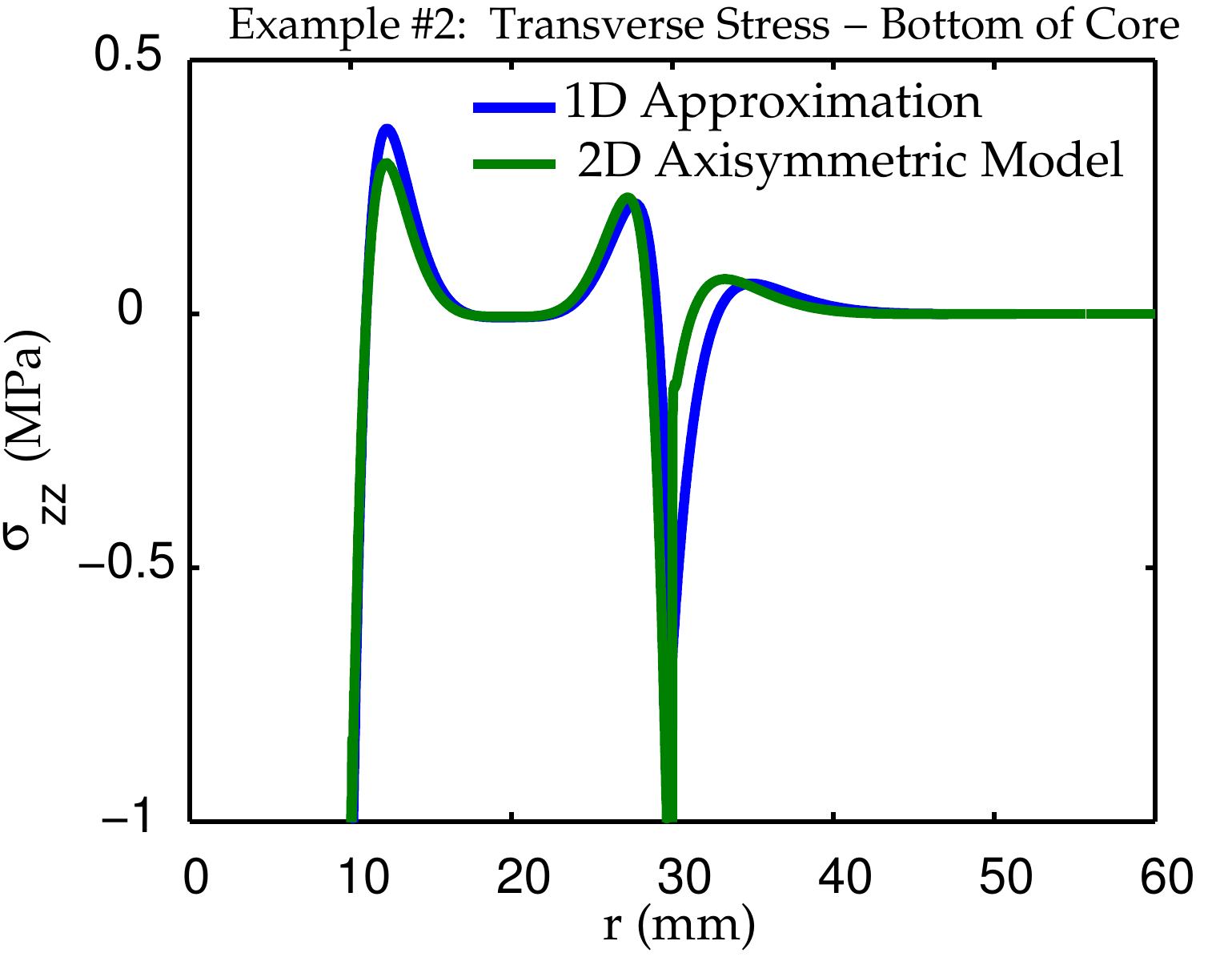}}\\
      (a) Transverse stress ($\sigma_{zz}$)- core bottom.
    \end{minipage}
    \begin{minipage}[b]{0.5\linewidth}
      \centering
      \scalebox{0.45}{\includegraphics{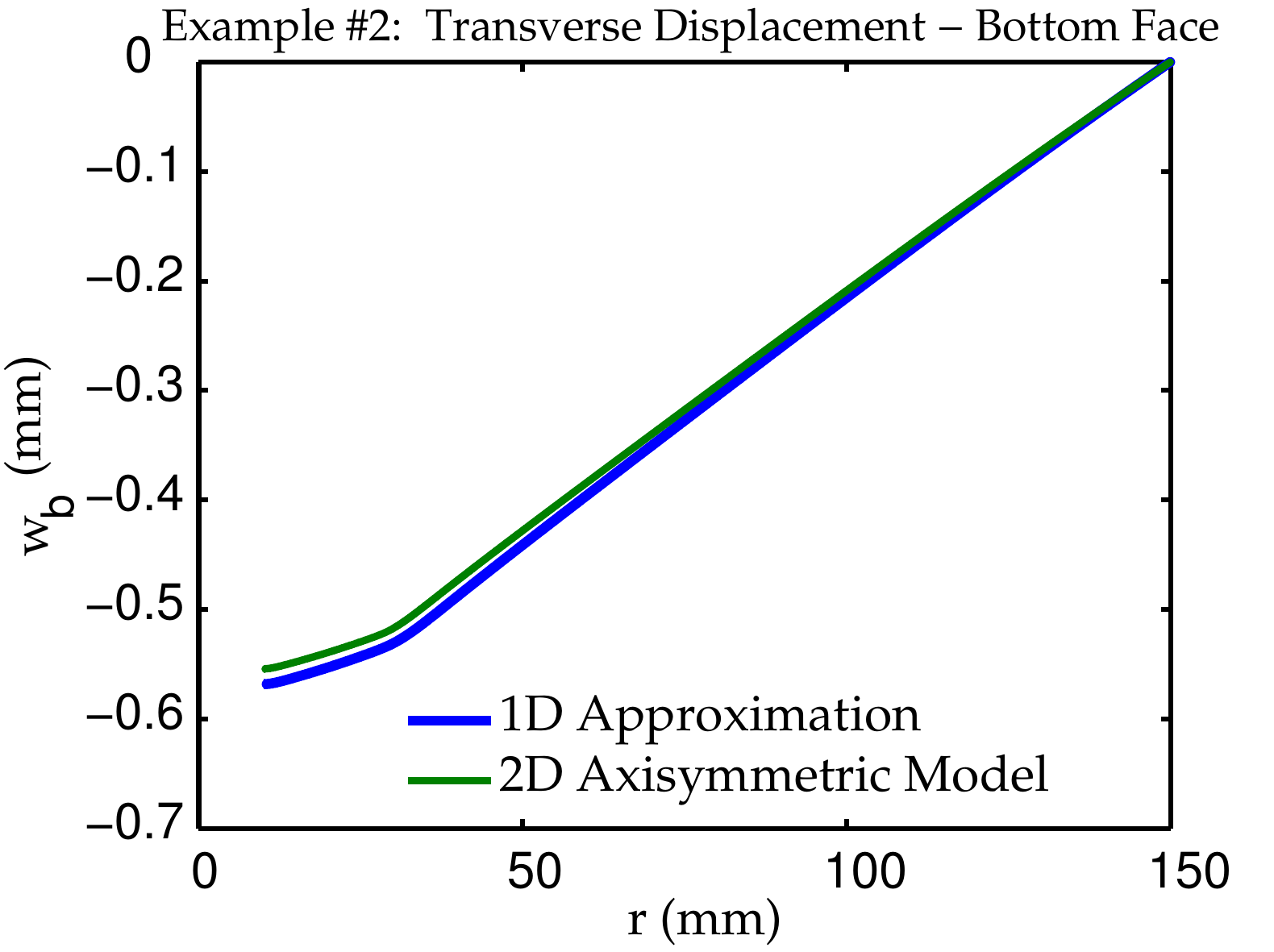}}\\
      (b) Transverse disp. ($w_b$)- core bottom.
    \end{minipage}
    \caption{Comparison of stresses and displacements at the bottom of the 
      core from one-dimensional and two-dimensional
      finite element simulations for the model in Table~\ref{tab:Ex2}.}
    \label{fig:szb-Ex2}
  \end{figure}

  The second example problem is taken from \cite{Thomsen98a}, with the
  parameters given in Table \ref{tab:Ex2}.
  Once again, figure \ref{fig:w-Ex2}(a) shows the out of plane displacements given
  by the sandwich theory and the axisymmetric simulations, and figure
  \ref{fig:w-Ex2}(b) gives the radial displacements.  As in the first example, the
  results match reasonably well, suggesting that the sandwich theory captures
  the important physics of the problem.  The stresses shown in Figure~\ref{fig:srz-Ex2}
  also show that the one- and two-dimensional models predict similar results.  The
  values of transverse stress and displacement at the bottom of the core are
  shown in Figure~\ref{fig:szb-Ex2}.
  \begin{table}[htb!]
    \begin{center}
      \caption{Geometric and material parameters for Example 2} \label{tab:Ex2}
      \begin{tabular}{lllllll}
      Geometry (mm): & $r_i=10.0$& $r_p=30.0$& $r_a=150.0$ & $c=5.0$ 
        & $f^{\Ttop}=0.5$ & $f^{\Tbot}=0.1$ \\
      Face Sheets (GPa) : & $E_1=40.0$ & $G_1=14.8$ \\
      Potting (GPa):  &  $E_p=2.5$ & $G_p=0.93$ \\
      Honeycomb (MPa): & $E_h=310$ & $G_h=138$
      \end{tabular}
    \end{center}
  \end{table}

\section{Summary and Conclusions}
A detailed on-dimensional theory for sandwich panels with inserts has been derived.  The
approach follows that used by Thomsen~\cite{Thomsen98a}.  The models has been discretized using
a finite element approach.  The one-dimensional model produces results that are close to those of 
a two-dimensional axisysmmetric finite element model.  Both models assume that the core is homogeneous, 
indicating that the one-dimensional model might be well suited for small deformations of sandwich 
specimens with foam cores.  Further work is need to find nonlinear one-dimensional models of
sandwich panels with inserts.

%
%
\section*{Acknowledgments}
This work was funded by a subcontract of the University of Auckland Research for Industry grant 
``Advanced Composite Structures'' from the New Zealand Foundation for Research, Science, and Technology.

\glsaddall
\printglossary

\bibliographystyle{unsrt}
\bibliography{../mybiblio}

\begin{thebibliography}{10}

\bibitem{Plantema66}
F.~J. Plantema.
\newblock {\em Sandwich Construction}.
\newblock John Wiley and Sons, New York, 1966.

\bibitem{Burton97}
W.~S. Burton and A.~K. Noor.
\newblock Assessment of continuum models for sandwich panel honeycomb cores.
\newblock {\em Comput. Methods Apl. Mech. Engrg.}, 145:341--360, 1997.

\bibitem{VuQuoc97}
L.~Vu-Quoc, I.~K. Ebcio\u{g}lu, and H.~Deng.
\newblock Dynamic formulation for geometrically-exact sandwich shells.
\newblock {\em Int. J. Solids Struct.}, 34(20):2517--2548, 1997.

\bibitem{Anderson98}
T.~Anderson, E.~Madenci, W.~S. Burton, and J.~Fish.
\newblock Analytical solution of finite-geometry composite panels under
  transient surface loading.
\newblock {\em Int. J. Solids Struct.}, 35(12):1219--1239, 1998.

\bibitem{Barut01}
A.~Barut, E.~Madenci, J.~Heinrich, and A.~Tessler.
\newblock Analysis of thick sandwich construction by a \{3,2\}-order theory.
\newblock {\em Int. J. Solids Struct.}, 38:6063--6077, 2001.

\bibitem{Barut02}
A.~Barut, E.~Madenci, T.~Anderson, and A.~Tessler.
\newblock Equivalent single-layer theory for a complete stress field in
  sandwich panels under arbitrary distributed loading.
\newblock {\em Composite Structures}, 58:483--495, 2002.

\bibitem{Thomsen98}
O.~T. Thomsen and W.~Rits.
\newblock Analysis and design of sandwich plates with inserts - a higher order
  sandwich plate theory.
\newblock {\em Composites Part B}, 29B:795--807, 1998.

\bibitem{Thomsen98a}
O.~T. Thomsen.
\newblock Sandwich plates with 'through-the-thickness' and 'fully potted'
  inserts: evaluation of differences in structural performance.
\newblock {\em Composites Structures}, 40(2):159--174, 1998.

\bibitem{Thomsen00}
O.~T. Thomsen.
\newblock High-order theory for the analysis of multi-layer plate assemblies
  and its application for the analysis of sandwich panels with terminating
  plies.
\newblock {\em Composites Structures}, 50:227--238, 2000.

\bibitem{Rabin02}
O.~Rabinovitch and Y.~Frostig.
\newblock High-order behavior of fully bonded and delaminated circular sandwich
  plates with laminated face sheets and a ``soft'' core.
\newblock {\em Int. J. Solids Struct.}, 39:3057--3077, 2002.

\bibitem{Kulikov08}
G.~M. Kulikov and E.~Carrera.
\newblock Finite deformation higher-order shell models and rigid-body motions.
\newblock {\em Int. J. Solids Struct.}, 45:3153--3172, 2008.

\bibitem{VuQuoc00}
L.~Vu-Quoc and I.~K. Ebcio\u{g}lu.
\newblock General multilayer geometrically-exact beams/1-d plates with
  deformable layer thickness: Equations of motion.
\newblock {\em Z. Angew. Math. Mech.}, 80:113--136, 2000.

\bibitem{Frostig05}
Y.~Frostig, O.~T. Thomsen, and I.~Sheinman.
\newblock On the non-linear high-order theory of unidirectional sandwich panels
  with a transversely flexible core.
\newblock {\em Int. J. Solids Struct.}, 42:1443--1463, 2005.

\bibitem{Arciniega07}
R.~A. Arciniega and J.~N. Reddy.
\newblock Large deformation analysis of functionally graded shells.
\newblock {\em Int. J. Solids Struct.}, 44:2036--2052, 2007.

\bibitem{Arciniega07a}
R.~A. Arciniega and J.~N. Reddy.
\newblock Tensor-based finite element formulation for geometrically nonlinear
  analysis of shell structures.
\newblock {\em Comput. Methods Appl. Mech. Engrg.}, 196:1048--1073, 2007.

\bibitem{Hohe08}
J.~Hohe and L.~Librescu.
\newblock Recent results on the effect of the transverse core compressibility
  on the static and dynamic response of sandwich structures.
\newblock {\em Composites: Part B}, 39:108--119, 2008.

\end{thebibliography}

\end{document}